\documentclass[final,5p,times,onecolumn,authoryear]{elsarticle}
\usepackage[backref]{hyperref}
\usepackage[utf8]{inputenc}
\usepackage{graphicx}
\usepackage[normalem]{ulem}
\usepackage{textcomp}
\usepackage{natbib}
\usepackage{color}
\setcitestyle{authoryear,open={(},close={)}} 
\usepackage{amsmath}	
\usepackage{amssymb}	
\usepackage{lipsum}
\usepackage{tabularx}
\usepackage{epsfig}
\usepackage{xcolor}
\usepackage{lscape, longtable}	
\hypersetup{colorlinks,linkcolor={blue},citecolor={blue},urlcolor={red}}
\journal{New Astronomy}
\definecolor{midnightblue}{rgb}{0.1, 0.1, 0.44}
\definecolor{pinegreen}{rgb}{0.0, 0.47, 0.44}
\definecolor{persianplum}{rgb}{0.44, 0.11, 0.11}
\definecolor{purpleheart}{rgb}{0.41, 0.21, 0.61}
\definecolor{byzantium}{rgb}{0.44, 0.16, 0.39}
\definecolor{brightmaroon}{rgb}{0.76, 0.13, 0.28}
\definecolor{apricot}{rgb}{0.98, 0.81, 0.69}
\definecolor{dpbrown}{RGB}{153,0,0}
\definecolor{dpgreen}{RGB}{51,102,0}
\definecolor{violet}{RGB}{51,51,255}
\definecolor{forest_green}{RGB}{34,139,34}

\begin{document}

\begin{frontmatter}



\title{Thermal Conduction and {\color{black}{Thermal-Driven}} Winds in Magnetized Viscous Accretion Disk Dynamics}



\author{Biplob Sarkar$^{1*}$,
Indu Kalpa Dihingia$^2$, Ranjeev Misra$^3$\\}

\address{$^{1*}$Department of Applied Sciences, Tezpur University, Napaam-784028, Tezpur, Assam, India\\
$^{2}$Tsung-Dao Lee Institute, Shanghai Jiao-Tong University, Shanghai, 520 Shengrong Road, 201210, People's Republic of China\\
$^{3}$Inter-University Center for Astronomy and Astrophysics, Ganeshkhind, Pune, Maharashtra 411007, India\\
$\qquad$\\
$^{*}$Corresponding author(s). E-mail(s): biplobs@tezu.ac.in
}




\begin{abstract}
This paper investigates the effects of saturated thermal conduction (TC) and {\color{black}{thermal-driven}} winds {\color{black}{(TDWs)}} on magnetized advection-dominated accretion onto a rotating black hole (BH). We incorporate dissipative processes in the magnetized accretion flow and expect the accretion disk to be threaded by predominantly toroidal and turbulent magnetic fields. We solve the magnetohydrodynamics equations and construct a self-consistent steady model of the magnetized accretion flow surrounding a rotating BH, which includes TC and {\color{black}{TDWs}}. We seek global accretion solutions spanning from the BH horizon to a large distance and analyze the solution's characteristics as a function of dissipation parameters. Accretion solutions with multiple critical points may exhibit shock waves if they meet the standing shock criteria. We found steady, global transonic, and shocked accretion solutions around the rotating BH. In particular, the wind parameter ($m$) and the saturated conduction parameter ($\Phi_{\rm s}$) significantly influence the dynamical behavior of shocks. The shock location moves away from the BH horizon as $\Phi_{\rm s}$ and $m$ increase, assuming fixed conditions at the disk's outer edge. Our formalism explains the declining phase of BH outbursts, characterized by a monotonic decrease in QPO frequency as the burst decays. Based on our findings, we conclude that the combined effect of $\Phi_{\rm s}$ and $m$ parameters substantially alters the steady shock specific energy vs angular momentum parameter space and also modifies the corresponding post-shock luminosity vs QPO frequency parameter space. We propose, based on our theoretical model, that the $\Phi_{\rm s}$ and $m$ parameters may significantly influence the evolution of the BH outbursts.

\end{abstract}


\end{frontmatter}

\onecolumn
\section{Introduction}
\label{introduction}

{\color{black}{Astrophysical observations have revealed accretion disks (ADs) around stellar-mass black holes (BHs) and supermassive BHs (SMBHs) \citep[\& references therein]{Abramowicz2013,Yuan-Narayan2014,C2015ARep}. As matter from a companion star is gravitationally pulled into the BH, it may also attract material from the intergalactic medium. The accreting gas becomes magnetized due to flux freezing, and to spiral inward, matter within the AD must lose angular momentum. The accretion flow (AF) around a BH must transition through a sonic point near the event horizon (EH), changing from subsonic (distant from the BH) to supersonic \citep{Liang1980ApJ,Abramowicz1990,Misra1996}. This concept, refined by various researchers, is crucial for understanding the spectral and timing behavior of radiation emitted from ADs \citep{Chakrabarti1989,Chakrabarti1996a,Mukhopadhyay2003,Ray2007CQGra,Mukhopadhyay2008,Das2018ASSP}.}}

{\color{black}{Magnetic fields play a crucial role in astrophysical processes of BH AFs \citep{Abramowicz2013}, as they facilitate the outward transport of angular momentum necessary for mass accretion. The rotational energy of the BH can drive relativistic jets through magnetic field lines \citep{Blandford-Znajek1977}. Additionally, Magneto-Rotational Instability (MRI) contributes to disk viscosity, inducing turbulent AFs \citep{Balbus-Hawley1991}. The dominant toroidal component of the magnetic field necessitates consideration of the appropriate magnetic induction equation \citep{Hirose2006,Machida2006}. Recent studies have investigated optically thin, advection-dominated ADs around BHs incorporating magnetic fields \citep{Oda2010,Oda2012}.}}

{\color{black}{\citet{Sarkar2016,Sarkar2018,Sarkar-Das2018JApA,Das-Sarkar2018} investigated the impact of viscosity, toroidal magnetic fields, and radiative processes on the dynamics of global AFs around BHs. Their studies revealed the formation of standing shocks in self-consistent global accretion solutions. Furthermore, \citet{Sarkar2018} demonstrated that magnetic fields significantly influence the AD's luminosity and emerging spectrum. More recently, \citet{Sarkar2020RAA} discovered a strong correlation between maximum shock luminosity and observed radio jet kinetic power in several stellar-mass BH sources.}}

{\color{black}{Shock waves arise in global accretion solutions around BHs due to the centrifugal potential barrier encountered by the AF near the BH \citep{Fukue1987PASJ,Chakrabarti1989,DCC2001,Sarkar2013ASInC,Das2014BASI,Aktar2015,Saha2016,Shilpa2020}. These shocks trigger abrupt, discontinuous changes in flow properties across two distinct regions of the fluid \citep{2007CC,DC2009,Sarkar2015ASInC,DDM2018,Sarkar2020RAA}. The conversion of kinetic to thermal energy causes the post-shock (PS) region to puff up \citep{2013Giri,M2017ApJ,G2018MNRAS,M2018ASSP}. Crucially, shocked global accretion solutions around BHs explain spectral behavior, luminosity variability, and quasi-periodic oscillations (QPOs) \citep{Molteni1999,Das_T_K2002,Okuda2007,OTM2008,Giri2015,G2018ASSP,N2018ASSP}.}}

{\color{black}{X-ray data are crucial for studying accreting BHs, with numerous reports on QPO observations and potential explanations \citep{2004A&AG,2008PASJM,2010exruS,Falanga2015pabh,M2018ASSP}. The oscillation of the shock or PS region in an AF is believed to cause low- and intermediate-frequency QPOs in BH candidates \citep{1996ApJM,CM2000ApJ,Giri2015,D2018ASSP}. Specifically, shock oscillation is likely if the Rankine-Hugoniot criterion is not met \citep{Ryu1997ApJ} or if there is a reasonable match between the infall and PS regions' cooling timescales \citep{1996ApJM,2015MNRASC}.}}

{\color{black}{Numerical simulations have investigated various aspects of AFs and outflows, particularly in the context of BHs \citep[e.g.][]{2021MNRAS_Dihingia}. \citet{2003aApJProga,2003bApJProga} carried out hydrodynamic and magnetohydrodynamic simulations of low-angular momentum AFs using 2D axisymmetric models, incorporating latitude-dependent angular momentum, rotation, and magnetic fields. Building on this work, \citet{2008ApJMP,2009MNRASMP} performed time-dependent simulations to explore the effects of adiabatic index, gas temperature, and magnetic fields on flow dynamics, jet formation, and disk thickness around BHs. \citet{Janiuk2009ApJ} extended this research by investigating low-angular momentum accretion using 3D simulations, initializing with a spherically symmetric Bondi-type matter distribution and introducing latitude-dependent angular momentum, although their study did not address the formation and stability of standing shocks.}}  

{\color{black}{BH AD winds are ubiquitous \citep{Bu2016,Y2016ASSL,Cui2020,Cui_Yuan2020}, originating from thermal driving due to heating of the external region by the hot internal region \citep{Begelman1983,Abramowicz2013}. Numerical simulations \citep{Yuan-F2012,Giri2012,Bu2016,Cui2020} and observational studies \citep{Wang2013,Done2018,Higginbottom2018,Tomaru2019} confirm the existence of thermal-driven winds (TDWs). Analytical studies have also investigated TDW generation from ADs around compact objects \citep{Takahara1989,Misra2001ApJ,Abbassi2010,Abbassi2012,Dubus2019}. Thermal driving is particularly significant in X-ray-irradiated ADs, such as X-ray binary systems \citep{Begelman1983,1996ApJW,Tomaru2023}. For TDWs to occur, gas must reach temperatures sufficient to overcome gravitational binding \citep{2000ApJProga}. Furthermore, \citet{1983ApJ271B} showed that hard X-ray irradiation of the outer disk creates a hot coronal gas layer, emphasizing the crucial role of winds in the accretion energy budget and potentially inducing instabilities in the AF. \citet{Lin2001} carried out theoretical investigation on the potential existence of TDWs in X-ray binary systems. Building on this concept, \citet{2002ApJP} investigated radiation-driven outflows from ADs, exploring mechanisms driven by electron and line interactions, providing valuable insights into the theoretical aspects and properties of AD winds.}}

{\color{black}{Thermal conduction (TC) plays a crucial role in ADs surrounding BHs, influencing heating, radiative processes, and disk structure \citep{Shadmehri2008,Ghasemnezhad2018,Meyer-Meyer2006}. TC transports thermal energy from the hot inner region to the cold outer region \citep{Ghasemnezhad2018}, impacting wind properties and spectral transitions in BH X-ray binaries \citep{Bu-Wu-Yuan2016,Nakamura2019}. Studies have shown TC's significant effects on AD dynamics, including ejection efficiency, matter flow, and jet launching \citep{Rezgui2019ApJ,Rezgui2022MNRAS}. Recent research has explored TC's influence on transonic properties in hot AD with low angular momentum \citep{Mitra2023}. This paper investigates energy transport via saturated TC in AD, following \citet{Cowie-McKee1977} and \citet{Tanaka-Menou2006}.}}

{\color{black}{Most BHs are rotating, with some rotating rapidly \citep{Shafee2006,Gou2009,Aschenbach2010,Liu2010,Gou2011,Ludlam2015}. Studying accretion around BHs requires a general relativistic treatment. Pseudo-Newtonian potentials (PNPs) effectively model the gravity of rotating BHs \citep{ABN1996,Bhattacharjee_et_al2022,CK1992,CM2006MNRAS,Dihingia2018c,Lu1995ChA,Mukhopadhyay2002,SK1999}. This paper utilizes the PNP by \citet{ABN1996} to incorporate general relativistic effects, acknowledging its limitations, discussed in the conclusion.}}

This paper aims to: (i) construct a self-consistent steady model incorporating shocks, TC, and {\color{black}{TDWs}} in the context of a rotating BH's AF (a task not attempted previously), (ii) investigate the characteristics of accretion solutions as a function of dissipation parameters and search for global solutions spanning a significant distance from the BH horizon, (iii) identify standing shocks in global magnetized accretion solutions and examine their properties in relation to TC and {\color{black}{TDWs}}, (iv) estimate PS disk luminosity and QPO frequencies associated with global magnetized shocked accretion solutions.

The paper is organised as follows. The governing equations and assumptions of our model are presented in Section 2. In Section 3, the numerical findings are provided in detail. In Section 4, we present a comprehensive analysis and summary of the current work.


\section{Assumptions for the Model, Governing Equations, and Solution Process}

Our study examines magnetized AF that is axisymmetric, thin, and steady around a rotating BH. The universal gravitational constant, $G$, the BH’s mass, $M_{\rm BH}$, and light’s speed, $c$, are used to represent the radial coordinate ($x$) in units of $r_g = GM_{\rm BH}/c^2$ throughout the paper. In addition, the flow velocity and time units are $c$ and $GM_{\rm BH}/c^3$, respectively. The coordinate system we consider is ($x$, $\phi$, $z$), with the BH situated in the coordinate system’s centre and the AF contained in the $x-\phi$ plane. 

The dynamic flow variables, which include sound speed, radial velocity, specific angular momentum, mass density, adiabatic index, temperature, gas pressure, magnetic pressure, specific entropy, radial component of magnetic field, and azimuthal component of magnetic field are represented by the following symbols: $c_s$, $u$, $\lambda$, $\rho$, $\gamma$, $T$, $p_{\rm gas}$, $p_{\rm mag}$, $s$, $B_{x}$, and $B_{\phi}$, respectively. Therefore, the equations for the conservation of mass flux \citep{Blandford-Begelman1999, Shadmehri2008, Abbassi2012,Y2016ASSL}, radial momentum, azimuthal momentum, induction equation, and entropy generation are, respectively,
$$
\dot{M}=2\pi u\Sigma x = \dot{M}_{\rm out}\left(\frac{x}{x_{\rm out}}\right)^m,
\eqno(1)
$$
$$
{u\frac{\displaystyle du}{\displaystyle dx} + \frac{\displaystyle 1}{\displaystyle \rho}\frac{\displaystyle dp_{\rm tot}}{\displaystyle dx} 
- \frac{\displaystyle \lambda^2}{\displaystyle x^3} + F(x) + \frac{\displaystyle \left<B_{\phi} ^2\right>}{\displaystyle 4\pi x \rho} = 0},
\eqno(2)
$$
$$
{u \Sigma x}\frac{\displaystyle d\lambda}{\displaystyle dx}+ \frac{\displaystyle d}{\displaystyle dx}(x^2\mathbb{T}_{x\phi}) = 0,
\eqno(3)
$$
$$
{\bf \nabla} \times
\left({\vec{u}} \times <B_{\phi}>\hat{\phi} -{\frac{4\pi}{c}}\eta {\vec{j}}\right) = 0,
\eqno(4)
$$
$$
u T\Sigma \frac {ds}{dx}=\frac{Hu}{1 - \gamma}
\left(\frac{\displaystyle \gamma p_{\rm gas}}{\displaystyle \rho}\frac{\displaystyle d\rho}{\displaystyle dx} - \frac{\displaystyle dp_{\rm gas}}{ dx}\right)=\mathcal{Q}^- - \mathcal{Q}^+ - \mathcal{Q}_{\rm cond}.
\eqno(5)
$$

In principle, equations (2-4) should also be modified in the presence of mass outflow (see, for example, \citet{Xie2008,Bu2009}). Nonetheless, as pointed out by \citet{Yuan-Narayan2014}, the density profile is most likely where an outflow has its greatest impact. Therefore, to obtain a good initial approximation, we leave equations (2-4) exactly as given following the work by \citet{Yuan-Narayan2014}.

When following the pseudo-Newtonian approach of \citet{ABN1996,Saha2016,AND2019}, the pseudo-Kerr force's expression is,

$$
F(x) = \frac{\displaystyle 1}{\displaystyle x^{2-\delta}(x - x_{\mathcal{H}})^{\delta}}
\eqno(6)
$$
where the radial coordinate is indicated by $x$, and the EH's location is indicated by $x_{\mathcal{H}}$. From general theory of relativity, the EH is precisely expressed as $x_{\mathcal{H}} = 1 + \sqrt{(1 - a_{\rm BH}^2)}$. The exponent $\delta$ is given as $\delta = \frac{\displaystyle x_{\mathcal{MS}}}{\displaystyle x_{\mathcal{H}}} -1$, where $x_{\mathcal{MS}}$ represents the marginally stable circular orbit's radius \citep{Bardeen1972,ST1983,NF1989}. We find the equation for the marginally stable circular orbit's radius as follows \citep{Bardeen1972,ST1983}:
$$
x_{\mathcal{MS}} = {\mathcal{Z}}_2 + 3 \mp {\sqrt{({\mathcal{Z}}_1 + 2{\mathcal{Z}}_2 + 3)(3 - {\mathcal{Z}}_1)}},
\eqno(7)
$$
where, ${\mathcal{Z}}_1 = (1-a_{\rm BH}^2)^{1/3}\left[(1+a_{\rm BH})^{1/3} + (1-a_{\rm BH})^{1/3}\right] + 1$, and ${\mathcal{Z}}_2 = ({\mathcal{Z}}_1^2 + 3a_{\rm BH}^2)^{1/2}$. Here, prograde and retrograde flow are indicated by the `$\mp$' symbol.
In this case, $a_{\rm BH}$ stands for the BH rotation parameter, which is the BH's specific spin angular momentum. An explicit representation of the corresponding BH potential is provided as \citep{WW2012,NSAD2017,MND2018}:
$$
\Psi_{\rm BH}(x) = \frac{1}{(\delta - 1)x_{\mathcal{H}}}\left[1 - \frac{x^{\delta -1}}{(x - x_{\mathcal{H}})^{\delta -1}}\right]
\eqno(8)
$$
for $x_{\mathcal{H}} < x$ and $\delta \neq 1$. The aforementioned pseudo-Kerr potential exactly corresponds to \citet{PW1980} potential for $a_{\rm BH} = 0$ and $\delta = 2$. Overall, this PNP potential agrees well with the results derived from Kerr geometry. Nevertheless, the accretion solutions differ within a range of 10\%-20\% for highly rotating BH with the results derived from Kerr metric \citep{ABN1996,Mondal2018ASSP}. 

In this case, $\Sigma$ denotes the flow's mass density averaged in the vertical direction, and $\dot{M}$ denotes the mass accretion rate \citep{Mat-et-al1984}. One may note that we always consider the inward radial velocity's direction to be positive in the current study. Moreover, $p_{\rm tot}$ represents the flow's total pressure, which is calculated as follows: $p_{\rm tot} = p_{\rm gas} + p_{\rm mag}$. The gas pressure in the disk is calculated using the ideal gas law, where $p_{\rm gas} = R\rho T/\mu$, with $R$ representing the gas constant and $\mu$ being the mean molecular weight. We examine a completely ionized hydrogen gas with $\mu = 0.5$. Also, $p_{\rm mag} = <B_{\phi}^2>/8\pi$ is the flow's magnetic pressure. Our definition of the flow's magnetic parameter is $\beta = p_{\rm gas}/p_{\rm mag}$. The definition of the magnetic parameter is used to determine the flow's total pressure, which is $p_{\rm tot} = p_{\rm gas} (1 + 1/\beta)$.  Additionally, $\mathbb{T}_{x\phi}$ represents the Maxwell stress's $x\phi$ component, which is the dominant component in the flow's vertically integrated total stress. When the AF's velocity in the radial direction is considerable, $\mathbb{T}_{x\phi}$ is obtained as
$$
\mathbb{T}_{x\phi} = \frac{<B_{x}B_{\phi}>}{4\pi}H = -\alpha_{T}(W + \Sigma u^2),
\eqno(9)
$$
where the disk half-thickness, the flow's vertically integrated pressure, and the constant of proportionality are represented by the symbols $H$, $W$, and $\alpha_T$, respectively. We treat $\alpha_T$ as constant throughout the flow, in keeping with the pioneering work of \cite{SS1973}. Equation (9) restores to the `$\alpha$-model' \citep{SS1973} when $u$ is negligible, as it would be for a Keplerian flow.

We determine the disk's half-thickness ($H$) as, $H(x)$ = $c_s x^{1/2} \gamma^{-1/2} F^{-1/2}$, assuming vertical hydrostatic equilibrium \citep{RM2010a,RM2010b,AND2019}. The adiabatic sound speed in this case is found to be $c_s=\sqrt {\gamma p_{\rm tot}/\rho}$. For the sake of the ensuing analysis, we take $\gamma$ to be constant throughout the flow and fix $\gamma=4/3$ (ultra-relativistic flow), in this study.

According to \citet{Blandford-Begelman1999}, $m$ (the wind parameter) is a constant with order unity in equation (1). $\mathcal{Q}^-$ and $\mathcal{Q}^+$, respectively, stand for the cooling and heating rates of the flow in equation (5), which is the equation for entropy generation. According to \cite{Machida2006,Hirose2006,KHB2007}, 3D MHD simulations have shown that the magnetic energy dissipated through the process of magnetic reconnection is the main process that contributes to the disk's heating. As a result, the heating rate is determined by,
$$
\mathcal{Q}^{+} = \frac{<B_{x}B_{\phi}>}{4\pi} x H \frac{d\Omega}{dx} = 
-\alpha_{T}(W + \Sigma u^2) x \frac{d\Omega}{dx},
\eqno(10)
$$
where $\Omega$ denotes the flow's angular velocity and `$<>$' signifies the azimuthal average. Moreover, we estimate the cooling rate $\mathcal{Q}^{-}$ using the relation $\mathcal{Q}^{-} = (1 - f_c)\mathcal{Q}^{+}$. Here, we include a parametric cooling factor $f_c$, whose value falls between the interval $0 \leq f_c \leq 1$, by the findings of \cite{Nar-Yi-1994,Nar-Yi-1995,Yuan-Narayan2014,Aktar2017MNRAS}. The AF exhibits advection dominance when $f_c = 1$, whereas cooling dominance occurs when $f_c = 0$. This paper fixes $f_c = 0.8$ for the following analysis.

In equation (5), $\mathcal{Q}_{\rm cond}$ represents the energy transfer through the process of TC. In hot AFs, the mean free path of electrons is much larger than the AF's length scale, making classical TC theory inapplicable. Hence, based on the work of \citet{Cowie-McKee1977,Tanaka-Menou2006} and \citet{2008MNRASA}, the heat flux is described by saturated conduction (SC) and $\mathcal{Q}_{\rm cond}$ is estimated as 
$$
\mathcal{Q}_{\rm cond} = -\frac{\displaystyle 2H}{\displaystyle x}\frac{\displaystyle d}{\displaystyle dx}(x\mathcal{F}_s).
\eqno(11)
$$ 
Here $\mathcal{F}_s$ represents the SC flux and is given by $\mathcal{F}_s = 5\Phi_{\rm s} \rho c_s^3$, where $\Phi_{\rm s}$ (called the SC parameter) is a constant less than or equal to 1. 

It has been reported in several studies that plasma must be considered to be two-temperature in the inner region of AD due to its high temperature \citep{Nar-Yi-1995,Lin2001,2005A&AM,RM2010b,2019JPhCS1336a2019S,Dihingia2020,Shilpa2020,2022Shilpa}. In this scenario, the electrons inside the flow may experience energy loss due to the inverse Comptonization process and also due to bremsstrahlung and synchrotron emission. Nonetheless, we've assumed a single temperature for the plasma in this study. As a result, we've neglected the inverse Comptonization of electrons, which requires a more complex two-temperature model. Additionally, we've ignored any interactions between ions and electrons, and instead estimated the electron temperature ($T_e$) using the simplifying relation $T_e = (\sqrt{m_e/m_p})T_p$, as proposed by \citet{CC2002}. In this case, $T_p$ denotes the proton temperature, and $m_p$ denotes the proton mass. It should be emphasised that, despite our assumption of a single temperature flow, at least in the inner disk region, it is anticipated that the temperature profiles of ions and electrons would diverge. \citet{MQ1997} showed that in the AF, the Coulomb collision-based energy exchange between ions and electrons is very inefficient. As a result, the ions and electrons are unable to reach thermal equilibrium. The heat obtained by viscous dissipation inside the disk is essentially retained by the ions. Further, the relativistic electrons lose energy through radiation more efficiently than the non-relativistic ions due to their significantly smaller mass. This leads to the formation of a two-temperature structure in the AF, where the electrons are cooler than the ions, particularly in the disk's inner region.

We now discuss equation (4), which is the induction equation and expresses the rate at which the magnetic field grows or escapes due to dynamo and diffusion. In place of the induction equation, we employ the rate of advection of the toroidal magnetic field defined as \citep{Oda-2007,Bu2009,FM2012MNRAS}
$$
\dot{\Phi} = \int u {B}_{\phi} dz = - (4\pi)^{1/2}u {B}_{0} (x) H,
\eqno(12)
$$
where,
\begin{eqnarray*}
{B}_{0} (x) && = \langle {B}_{\phi} \rangle \left(x; z = 0\right)  \nonumber \\
&& = \frac{(2^{5}{\pi})^{1/4}(R T \Sigma)^{1/2}}{(\mu H \beta)^{1/2}}
\end{eqnarray*}
represents the disk equatorial plane's mean azimuthal toroidal magnetic field \citep{Oda-2007}. If the dynamo and diffusion effects are disregarded for the steady-state AF under consideration, the quantity $\dot{\Phi}$ as mentioned above will remain constant. Nonetheless, the dynamo and diffusion effects may cause $\dot{\Phi}$ to vary with the radius in the AD in reality. We investigate the radial dependence of $\dot{\Phi}$ (equation 12) using the self-similar prescription for this quantity given by
$$
\dot{\Phi}\left(x\right) \equiv \dot{\Phi}_{\rm out}
\left(\frac{x_{\rm out}}{x} \right)^{\zeta},
\eqno(13)
$$
where, $\dot{\Phi}_{\rm out}$ is the rate of magnetic flux advection at the AD's outer edge, $x_{\rm out}$ and $\zeta$ is a constant. We choose $\zeta = 1$ throughout the paper, and thus, magnetic flux is enhanced as the AF approaches the BH. The radial dependence of $\dot{\Phi}$ (equation 13) is consistent with that prescribed by \citet{Machida2006,Oda-2007,FM2012MNRAS}.

\subsection{The solution process and transonic conditions}

The infalling matter is subsonic near the AD's outer edge ($x_{\rm out}$) and enters the BH supersonically because BH accretion processes are inherently transonic \citep{Chakrabarti1990,Chak1996b}. A critical point is typically characterized as the location in the radial direction where the flow undergoes a smooth transition from a subsonic state to a supersonic regime, marking a significant change in the flow's behavior. The critical point analysis is carried out as follows: solving equations (1)-(5), equation (9), and equations (12)-(13) simultaneously \citep[and references therein]{Sarkar2016}. It is provided by:

$$
\frac {du}{dx}=\frac{\mathcal{N}}{\mathcal{D}},
\eqno(14)
$$
where the numerator ($\mathcal{N}$) is calculated as,

$$
\mathcal{N} =\frac {2f_c\alpha^2_T I_n (m+1) (c_s^2g+\gamma u^2)^2}{\gamma^2 u x} - \frac {40\Phi_{\rm s} c_s^3}{x (1 + \beta)}-\frac {5 \Phi_{\rm s} (2m - 5) c_s^3}{x}
$$
$$
+\left[\frac{4f_c\alpha^2_T I_n g (c_s^2g+\gamma u^2)}{\gamma u} -\frac {u\{3+\beta(\gamma+1)\}}{(\gamma-1)(\beta + 1)}\right]
\left(\frac {\lambda^2}{x^3} - F\right)
$$
$$
+\left[- 5\Phi_{\rm s} c_s^3 + \frac {u c_s^2(2+ \gamma\beta)}{\gamma(\gamma-1)(\beta + 1)}-
\frac {2f_c\alpha^2_T c_s^2 I_n g(c_s^2g+\gamma u^2)}{\gamma^2 u}\right]
\left(\frac{1}{F}\frac{dF}{dx}\right)
$$
$$
+\frac {2c_s^2 u[3 + \beta(\gamma + 1)]}{\gamma (\gamma - 1)(\beta + 1)^2 x}+\frac {(2m-3)c_s^2 u(2\gamma\beta+3)}{2\gamma (\gamma - 1) x (1 + \beta)}
$$
$$
+\frac {2(3 -2m) f_c \alpha_T^2 I_n c_s^2 g(c_s^2g+\gamma u^2)}{\gamma^2 u x}-\frac {8 f_c\alpha_T^2 I_n c_s^2 g(c_s^2g+\gamma u^2)}{\gamma^2 (1 + \beta) u x}
$$
$$
- \frac {c_s^2 u (4\zeta +2m - 1)}{2\gamma (\gamma - 1)x (\beta + 1)}-\left(\frac {4f_c \alpha_T I_n \lambda}{\gamma x^2}\right) (c_s^2g+\gamma u^2) 
\eqno(15a)
$$

and the denominator ($\mathcal{D}$) is calculated as,
$$
\mathcal{D} = \frac {2c_s^2}{(\gamma-1)}\left[\frac{2}{\gamma(\beta + 1)} + \frac{\beta}{\beta + 1}\right] 
-\frac {\{3+(\gamma+1)\beta\}u^2}{(\gamma-1)(\beta + 1)}
$$
$$
+\frac{2f_c\alpha^2_T I_n}{\gamma} (c_s^2g+\gamma u^2)
\left[ (2g-1)-\frac {c_s^2g}{\gamma u^2}\right] 
$$
$$
+20\Phi_{\rm s}\gamma c_s u - \frac{10\Phi_{\rm s} c_s^3}{u}.
\eqno(15b)
$$
We define $g=I_{n+1}/I_{n}$ in the analysis above.

The derivative of $c_s$, $\lambda$, $\beta$, and $\dot{M}$ with respect to $x$ is then computed as,

$$
\frac{dc_s}{dx}= - \left( \frac{\gamma u}{c_s} - \frac{c_s}{u}\right)
\frac{du}{dx} - \frac{(2m-3)c_s}{2x} -\frac{c_s}{2F}\left(\frac{dF}{dx}\right) 
$$
$$
+ \frac{\gamma}{c_s}\left(\frac {\lambda^2}{x^3} - F\right) - \frac{2c_s}{x(1+\beta)}
\eqno(16)
$$

$$
\frac{d\lambda}{dx}=
\frac{\alpha_{T} x (u^2 \gamma - c_s^2g)}{u^2 \gamma}\frac{du}{dx}
+\frac{2 \alpha_{T} c_sxg }{\gamma u}\frac{dc_s}{dx}
$$
$$
+\frac{\alpha_{T}(m+1)(c_s^2g+\gamma u^2)}{\gamma u}
\eqno(17)
$$

$$
\frac{d\beta}{dx}= \left[\frac{4(\beta + 1)}{u}-\frac{3\gamma (\beta + 1)u}{c_s^2}\right]\frac{du}{dx}+\frac{3(\beta + 1)(3-2m)}{2x}
$$
$$
-\frac{2(\beta + 1)}{F}\left(\frac{dF}{dx}\right)+\frac{3\gamma(\beta + 1)}{c_s^2}\left(\frac{\lambda^2}{x^3}- F\right)
$$
$$
+\frac{(\beta + 1)(4\zeta+2m-1)}{2x} -\frac{6}{x}
\eqno(18)
$$

$$
\frac{d\dot{M}}{dx}=
\frac{m\dot{M}}{x}
\eqno(19)
$$

The infalling matter has to accrete onto the BH smoothly. This means that at every point along the flow, the flow variables must have finite values. Therefore, in order to preserve a smooth solution, if the denominator $\mathcal{D}$ in equation (14) is equal to zero at any radial position, the numerator $\mathcal{N}$ must also concurrently vanish there (${du}/{dx}= {0}/{0}$). A critical point ($x_c$) is a radial position of this kind. Thus, we have $\mathcal{N} = 0$ and $\mathcal{D} = 0$ as our two critical point conditions. With the condition that $\mathcal{D} = 0$, the expression of the Mach number ($M = u/c_s$) at $x_c$ is found as,
$$
\tau_1 M_c^4 + \tau_2 M_c^3+ \tau_3 M_c^2 + \tau_4 M_c + \tau_5 = 0,
\eqno(20)
$$
where
$$ 
\tau_1=2f_c\alpha^2_{T} I_n \gamma^2 (1+\beta_c)(1 - \gamma)(1 - 2g) - \{(1+\gamma)\beta_c + 3\}\gamma^2,
$$
$$
\tau_2= 20\Phi_{\rm s} \gamma^3 (\gamma - 1)(1+ \beta_c),
$$ 
$$
\tau_3=2\gamma(\gamma\beta_c + 2) + (2f_c\alpha^2_{T} I_n \gamma g)(1 - \gamma)(1 + \beta_c) (1-2g) + (2f_c\alpha^2_{T} I_n\gamma g) (1-\gamma)(\beta_c + 1),
$$
$$
\tau_4=10\Phi_{\rm s} \gamma^2 (1 - \gamma)(1 + \beta_c),
$$
$$
\tau_5=-2f_c\alpha^2_{T} I_n g^2 (1+ \beta_c)(\gamma - 1),
$$
where the definition of the parameter $g$ is $g = I_{n + 1}/I_n$. The sound speed ($c_{\rm sc}$) at $x_c$ may be obtained by applying the condition $\mathcal{N}=0$ to obtain a cubic equation of the following form:

$$
{\mathcal A_1}c_{\rm sc}^2 + {\mathcal A_2}c_{\rm sc} +{\mathcal A_3}= 0 ,
\eqno(21)
$$
where

$$
{\mathcal A_1} =  \frac {2f_c\alpha^2_T I_n (m + 1) (\gamma M_c^2 + g)^2}
{\gamma^2 x_c}$$
$$+\left[\frac{M_c^2(2+\gamma\beta_c)}{\gamma(\gamma - 1)(1 + \beta_c)} - 5\Phi_{\rm s} M_c\right]\left(\frac{1}{F}\frac{dF}{dx}\right) 
$$
$$
-\frac {2f_c\alpha^2_T I_n g (g+\gamma M_c^2)}
{\gamma^2}\left(\frac{1}{F}\frac{dF}{dx}\right)
$$
$$
 + \frac{2M_c^2\{\beta_c(1+\gamma) + 3\}}{\gamma(\gamma - 1)(1+\beta_c)^2 x_c} 
$$
$$
 + \frac{(2m -3)M_c^2(2\gamma\beta_c + 3)}{2(1+\beta_c)\gamma(\gamma - 1)x_c}-\frac {2(2m-3)f_c\alpha^2_T I_n g(\gamma M_c^2 + g)}{ \gamma^2 x_c}
$$
$$
-\frac {8f_c\alpha^2_T I_n g(\gamma M_c^2 + g)}{\gamma^2 (1+\beta_c)x_c}+ \frac{M_c^2(4\zeta + 2m -1)}{2\gamma(1-\gamma)(1+\beta_c) x_c},
$$
$$
-\frac{40\Phi_{\rm s} M_c}{x_c (1+\beta_c)} - \frac{5\Phi_{\rm s}(2m-5)M_c}{x_c}
$$

$$
{\mathcal A_2} = -\frac {4f_c \alpha_T I_n {\lambda_c} M_c (g+\gamma M_c^2)}{ x_c^2 \gamma},
$$

$$
{\mathcal A_3} = \left[\left(\frac{4f_c\alpha^2_T I_n g}{\gamma}\right)(g + \gamma M_c^2) - \frac {M_c^2\{\beta_c(\gamma + 1)+3\}}{(1 + \beta_c)(\gamma-1)} \right] \left(\frac{\lambda_c^2}{x_c^3}- F\right).
$$

Here, we use the subscript `c' to indicate the values of the flow variables estimated at $x_c$. 

At $x_c$, we can get the sound speed value ($c_{\rm sc}$) by providing the AF parameters in equation (21). Equation (20) may then be used to get the value of $u_c$. To examine the properties of the critical points, the calculated values of $u_c$ and $c_{\rm sc}$ are provided in equation (14). Two specific values of $du/dx$ exist at $x_c$; one corresponds to the accretion solution and the other to wind. $x_c$ is referred to be a `saddle' (X) type critical point if both of the velocity gradient's values there are real and have the opposite sign \citep{Liang1980ApJ,Chakrabarti1990,CD2004,Kato2020}. Since a global accretion solution passes via this critical point alone, it has special relevance. In the following analysis, we only take into account the accretion solutions since our goal in this work is to investigate the magnetized AF around a BH.

\section{Results:}

We now discuss the results. In this work, we examine how the SC parameter ($\Phi_{\rm s}$) and the wind parameter ($m$) affect the character of the accretion solutions. We are looking for global accretion solutions that connect the BH horizon to a large distance. We achieve this by solving equations for several flow variables ($u$, $c_{\rm sc}$, $\lambda$, plasma $\beta$, and $\dot{m}$) simultaneously. We use known boundary values of these flow variables at a specific distance ($x$) from the BH as a starting point. Additionally, the values of $\Phi_{\rm s}$, $a_{\rm BH}$, and $\alpha_{\rm T}$ must be detailed to obtain the solution. Since BH accretion solutions require transonic flow, we supply boundary values at the critical point and integrate the equations inward to the BH horizon and outward to the disk's outer edge. Joining the two parts yields a complete global transonic accretion solution. The flow may have multiple critical points, including inner critical points near the EH (denoted as $x_{\rm in}$) and outer critical points farther away (denoted as $x_{\rm out}$) \citep[\& references therein]{Gu2004ChPhL}. In all the results presented in this paper, we have assumed a fixed value of $\alpha_{\rm T} = 0.013$ (a representative value) and $a_{\rm BH} = 0.5$ (corresponding to a moderately rotating BH), unless otherwise specified. {\color{black}{The Table \ref{table:model_pars} collects all the model parameters, indicating which ones are kept fixed and which are varied throughout the analysis.}}

\begin{table}[]
\centering
\caption{\label{table:model_pars}{\color{black}{Model Parameters.}}}
\begin{tabularx}{1.0\textwidth} {  
  | >{\centering\arraybackslash}X
  | >{\centering\arraybackslash}X 
  | >{\centering\arraybackslash}X
  | >{\centering\arraybackslash}X | }
 \hline
 {\color{black}{Parameter Symbol}} & {\color{black}{Parameter Name}} & {\color{black}{Fixed/ Variable}} & {\color{black}{Value}}\\
 \hline
 {\color{black}{$\gamma$}}  & {\color{black}{Adiabatic index}} & {\color{black}{fixed}} & {\color{black}{4/3}}  \\
\hline
 {\color{black}{$f_c$}}  & {\color{black}{Parametric cooling factor}} &  {\color{black}{fixed}}          &  {\color{black}{0.8}}  \\
 \hline
 {\color{black}{$\alpha_{\rm T}$}}  & {\color{black}{Viscosity parameter}} &  {\color{black}{fixed}}          &  {\color{black}{0.013}}  \\
 \hline
 {\color{black}{$a_{\rm BH}$}}  & {\color{black}{BH rotation parameter}} &  {\color{black}{fixed}}          &  {\color{black}{0.5}}  \\
 \hline
 {\color{black}{$\mu$}}  & {\color{black}{Mean molecular weight}} &  {\color{black}{fixed}}          &  {\color{black}{0.5}}  \\
 \hline
  {\color{black}{$M_{\rm BH}$}}  & {\color{black}{Mass of the BH}} &  {\color{black}{fixed}}          &  {\color{black}{10 M$_\odot$}}  \\
 \hline
 {\color{black}{$\beta$}}  & {\color{black}{Plasma-$\beta$ or flow’s magnetic parameter}} &  {\color{black}{variable}}       &  {\color{black}{$--$}}  \\
 \hline
 {\color{black}{$\Phi_s$}}  & {\color{black}{Saturated conduction parameter}} &  {\color{black}{variable}}       &  {\color{black}{$--$}}  \\
 \hline
 {\color{black}{$m$}}  & {\color{black}{Wind parameter or mass-loss parameter}} &  {\color{black}{variable}}       &  {\color{black}{$--$}}  \\
 \hline
 {\color{black}{$\varepsilon$}}  & {\color{black}{Specific energy}} &  {\color{black}{variable}}       &  {\color{black}{$--$}}  \\
 \hline
 {\color{black}{$\lambda$}}  & {\color{black}{Specific angular momentum}} &  {\color{black}{variable}}       &  {\color{black}{$--$}}  \\
 \hline  
\end{tabularx}
\end{table}

\subsection{Global accretion solution without shock and the existence of outer critical point}

In Fig.~\ref{fig_1}(a), we considered the AD's outer edge at $x_{\rm edge} = 2000$. At the outer edge, we have fixed the plasma $\beta$ value to $\beta_{\rm edge} = 14430$, the specific angular momentum to $\lambda_{\rm edge} = 93.12$, and the accretion rate to $\dot{m}_{\rm edge}$ = 0.07314. We have also fixed the wind parameter at $m =$ 0.06. Then, we varied the SC parameter $\Phi_{\rm s}$. 

First, we consider the AF displayed using the green curve corresponding to $\Phi_{\rm s} = 3.35 \times 10^{-3}$, which starts subsonically from the outer edge with velocity $u_{\rm edge} = 6.53897\times 10^{-4}$ and sound speed $c_{\rm s,edge} = 5.78136\times 10^{-2}$. This subsonic flow approaches the BH horizon and passes through a critical point `I' close to it. Here, `I' denotes the location of the inner critical point located at $x_{\rm in}$ = 3.06796 where $\lambda_{\rm in}$ = 3.06900, $\beta_{\rm in}$ = 1.66536 and $\dot{m}_{\rm in}$ = 0.04958. After crossing point `I,' the flow becomes supersonic and enters the BH. 

Next, we decrease the SC parameter to $\Phi_{\rm s} = 3.0 \times 10^{-3}$ while keeping all other parameters fixed at the outer edge. We then adjust the velocity $u_{\rm edge} = 6.53157\times 10^{-4}$ and sound speed $c_{\rm s,edge} = 5.77970\times 10^{-2}$ at the outer edge (keeping other parameters fixed at $x_{\rm edge}$) to obtain the blue dashed curve, which also passes through the inner critical point located at $x_{\rm in}$ = 3.15703. 

This process is repeated with further reductions in the SC parameter. The red dotted curve corresponding to $\Phi_{\rm s} = 2.65 \times 10^{-3}$ also passes through the inner critical point located at $x_{\rm in}$ = 3.25468. However, an interesting behavior is observed as we continue to decrease $\Phi_{\rm s}$. For the given outer edge parameters, there exists a minimum value of $\Phi_{\rm s}$ (denoted as $\Phi_{\rm s}^{\rm min}$) for which the accretion solution passes through the inner critical point. For instance, the black solid curve, corresponding to $\Phi_{\rm s} = 2.30 \times 10^{-3}$ ($< \Phi_{\rm s}^{\rm min}$), fails to pass through the inner critical point. Instead, it passes through another critical point, `O,' farther away from the BH horizon. Here, `O' denotes the outer critical point's location positioned at $x_{\rm out}$ = 112.35425 where $\lambda_{\rm out}$ = 3.16920, $\beta_{\rm out}$ = 1043.6658 and $\dot{m}_{\rm out}$ = 0.06153. After crossing point `O', the flow becomes supersonic. We observe that the black solid curve has a closed solution topology, and this solution cannot connect the outer edge of the AD to the BH horizon. However, the important point to note in Fig.~\ref{fig_1} is the existence of the outer critical point `O' in addition to the inner critical point `I.' With regard to magnetized AFs including TC and {\color{black}{TDWs}}, the present paper is the first study on accretion solutions passing through an outer critical point. However, such solutions are of great interest since they are likely to exhibit shock waves supported by centrifugal repulsion.

Fig.~\ref{fig_1}(b) shows how the accretion rate varies radially for the different values of the SC parameter $\Phi_{\rm s}$ corresponding to the accretion solutions considered in Fig.~\ref{fig_1}(a). {\color{black}{Equation (1) of the paper reveals that the accretion rate ($\dot{m}$) follows a power-law relationship with the radial coordinate ($x$). Consequently, $\dot{m}$ does not directly depend on $\Phi_{\rm s}$, as also evident from equation (19). This explains the negligible dependence of $\dot{m}$ on $\Phi_{\rm s}$ observed in Fig.~\ref{fig_1}(b).}}

Fig.~\ref{fig_2}(a) shows how the plasma $\beta$ (logarithmic) parameter varies with the logarithmic radial coordinate $x$. The plasma $\beta$ parameter has the same value at the outer edge for all cases. However, it then follows different curves depending on the $\Phi_{\rm s}$. {\color{black} At a given radial coordinate, we observe that increasing the value of $\Phi_{\rm s}$ results in a decrease in the plasma $\beta$ value. Consequently, the flow becomes increasingly magnetically dominated as $\Phi_{\rm s}$ increases.}. {\color{black}{Fig.~\ref{fig_2}(b) displays the radial variation of the specific angular momentum ($\lambda$) profile, normalized by the Keplerian angular momentum $\lambda_{\rm Kep}$ = $x^{3/2}/(x - 2)$ \citep{Chak1996b,Das-Sarkar2018}. Equation (17) indicates that the derivative of specific angular momentum ($d\lambda/dx$) depends on $du/dx$, which is influenced by $\Phi_{\rm s}$ (see equation 14). Consequently, specific angular momentum depends on $\Phi_{\rm s}$, as evident from the four distinct lines in Fig.~\ref{fig_2}(b) showing variation relative to $\Phi_{\rm s}$. Moreover, at a fixed radial coordinate, our analysis shows that increasing the value of $\Phi_{\rm s}$ yields an increase in $\lambda$, which in turn indicates a decrease in the efficiency of outward angular momentum transport.}} The boundary conditions in Fig.~\ref{fig_2} are identical to those in Fig.~\ref{fig_1}(a).

\subsection{Global accretion solution including a shock}

\begin{figure}
	\centering 
	\includegraphics[height=8.0 cm,width=0.52\textwidth]{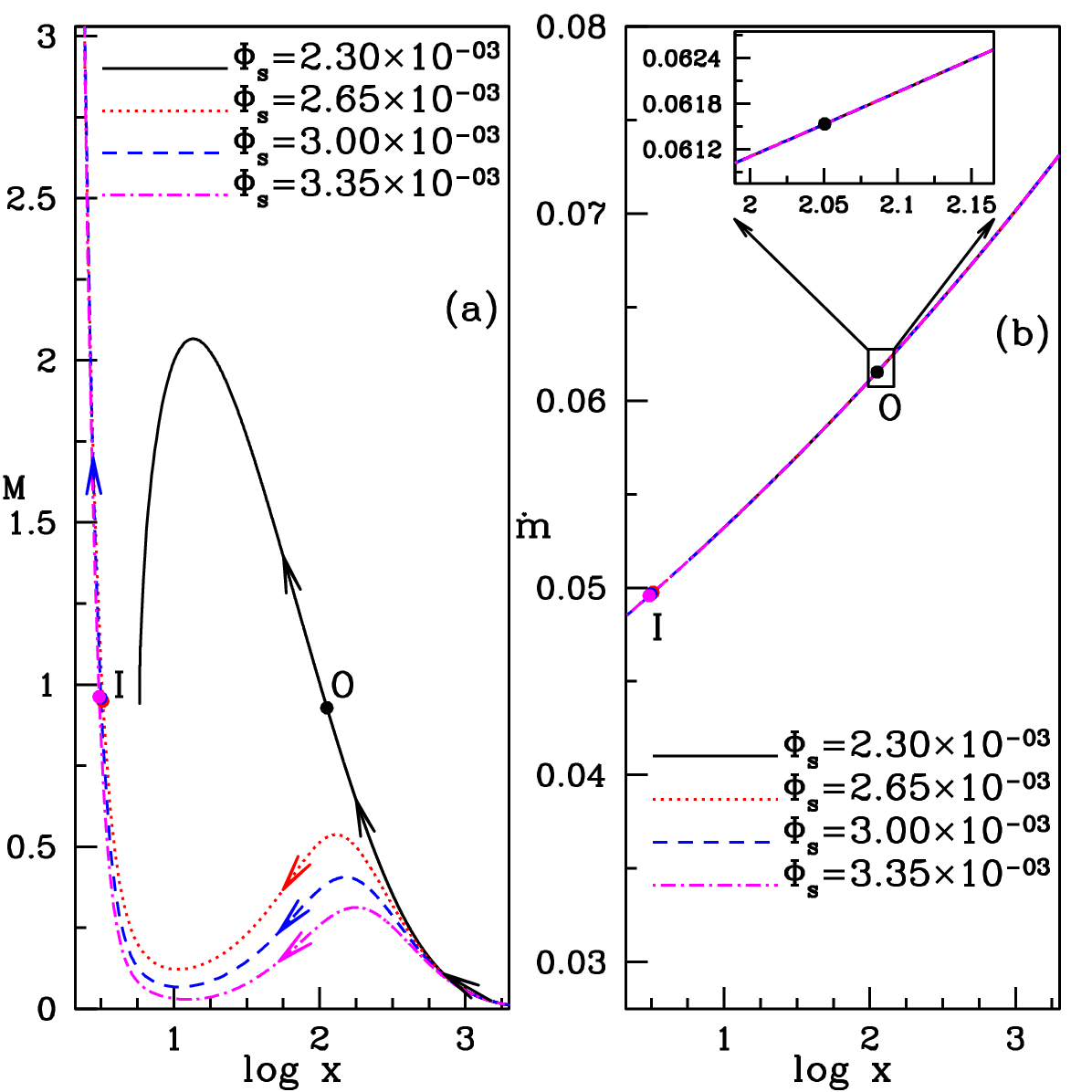}	
	\caption{(a) Radial dependence of Mach number ($M = u/a$) of the accreting matter for various values of the SC parameter ($\Phi_{\rm s}$). The AD's outer edge is located at $x_{\rm edge} = 2000$, where the values of $\beta$, $\lambda$, and $\dot{m}$ at the outer edge are 14430, 93.12, and 0.07314, respectively. The wind parameter is considered to be $m = 0.06$. The curves represent the results for different values of $\Phi_{\rm s}$: 3.35 $\times$ 10$^{-03}$ (dot-dashed), 3.00 $\times$ 10$^{-03}$ (dashed), and 2.65 $\times$ 10$^{-03}$ (dotted). A minimum value of the parameter $\Phi_{\rm s}$ (denoted as $\Phi_{\rm s}^{\rm min}$) is necessary for the accretion solution to pass through the inner critical point. If $\Phi_{\rm s}$ is less than $\Phi_{\rm s}^{\rm min}$, the solution only passes through the outer critical point (solid curve), as seen in the example where $\Phi_{\rm s}$ = 2.30 $\times$ 10$^{-03}$. The figure labels the inner critical point (I) and outer critical point (O), with arrows showing the direction of flow motion towards the BH. (b) The accretion rate ($\dot{m}$) decreases steadily as the flow approaches the BH horizon, due to mass loss, as shown in the plot of accretion rate versus radial distance, corresponding to the accretion solutions in panel (a). See the accompanying text for a detailed explanation.} 
	\label{fig_1}%
\end{figure}

\begin{figure}
	\centering 
	\includegraphics[height=8.0 cm,width=0.52\textwidth]{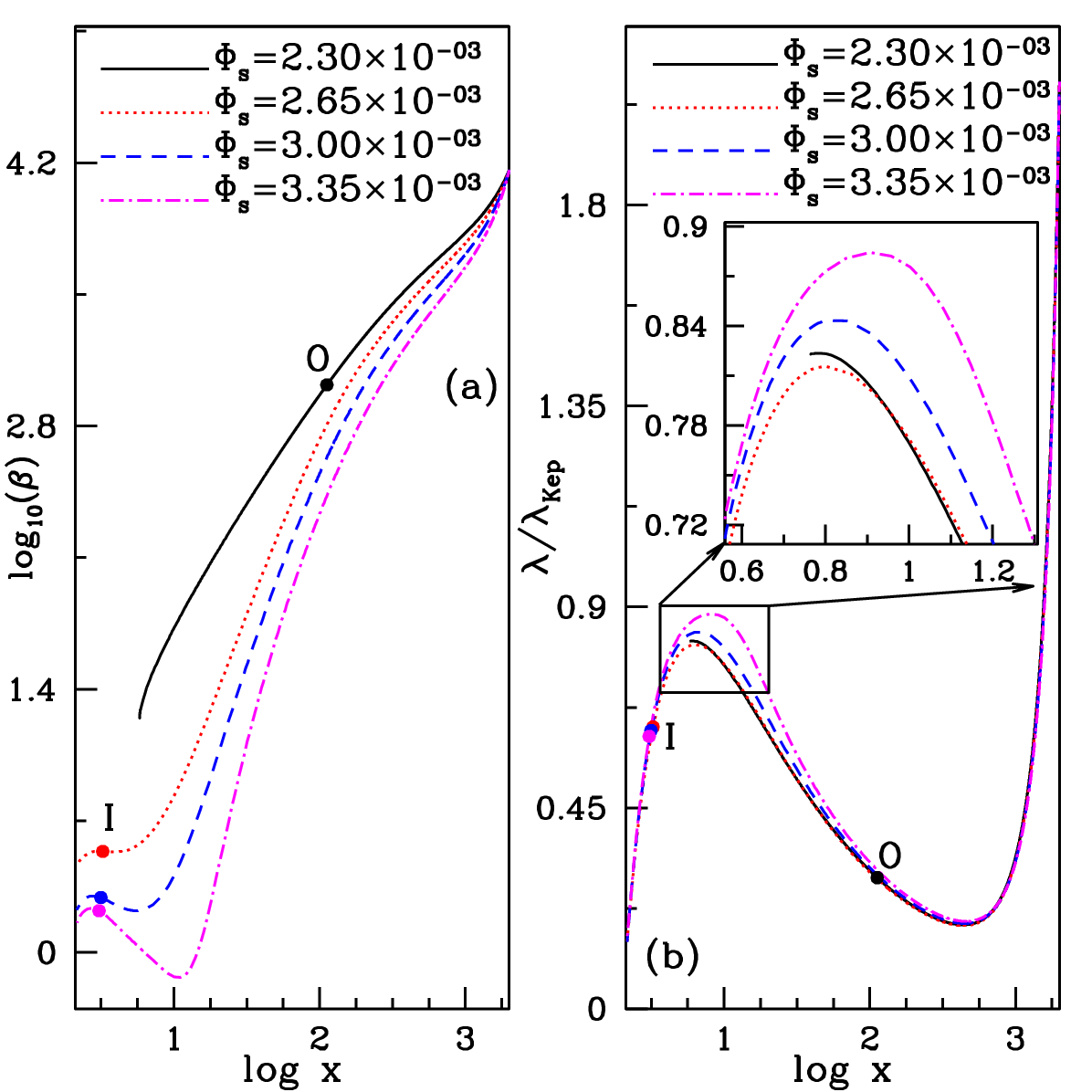}	
	\caption{Radial profile of plasma $\beta$ (logarithmic) (Panel a) and {\color{black}{normalized}} angular momentum ($\lambda$) (Panel b) of the AF for various $\Phi_{\rm s}$ values. Boundary conditions are identical to those in Fig.~\ref{fig_1}(a). Further details can be found in the text.} 
	\label{fig_2}%
\end{figure}
\begin{figure}
	\centering 
	\includegraphics[height=8.0 cm,width=0.52\textwidth]{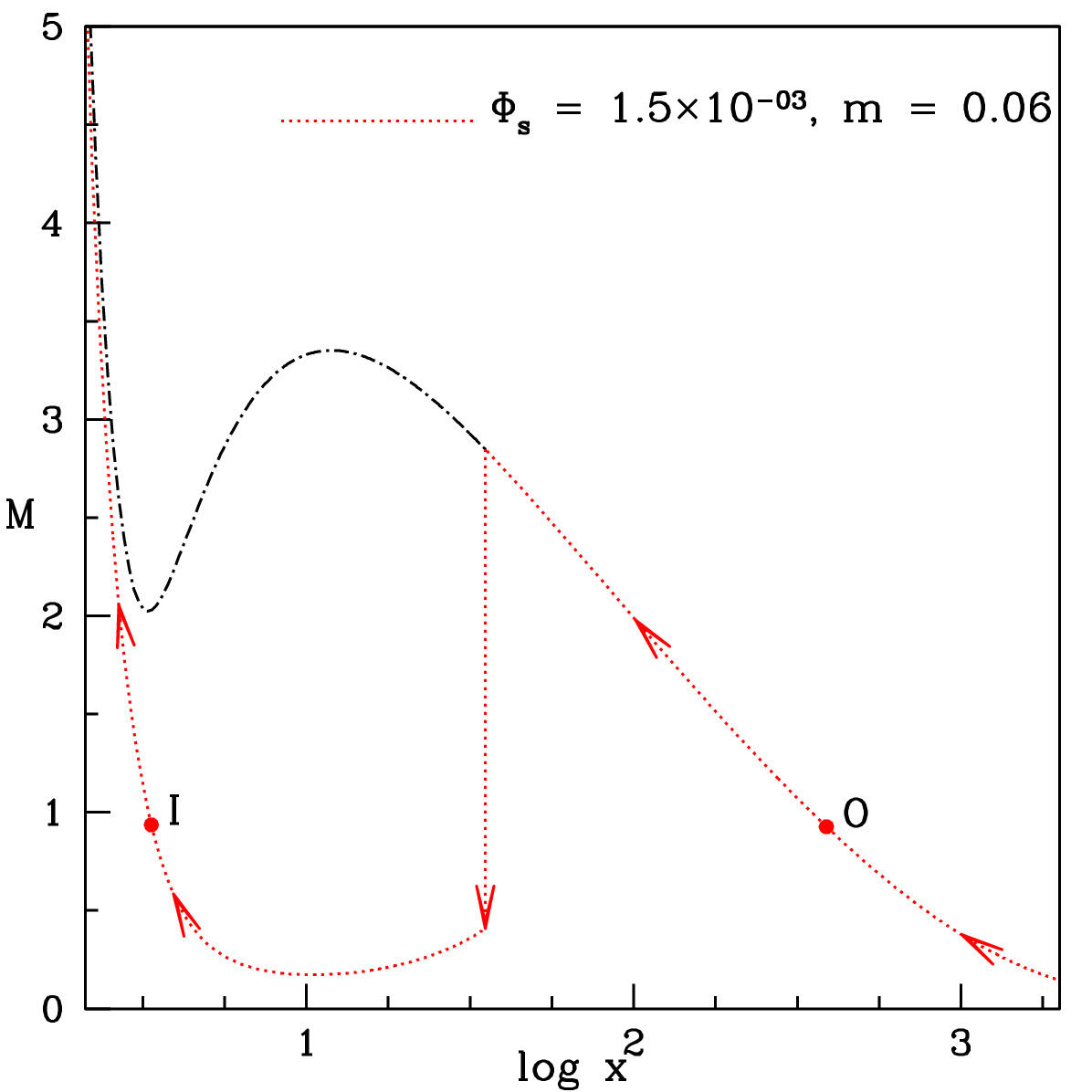}	
	\caption{A comprehensive global accretion solution is shown, featuring a shock wave located at $x_s$ = 35.22, as well as outer (O) and inner (I) critical points. The solution is characterized by specific parameters at the outer edge ($x_{\rm edge} = 2000$), including: a plasma-$\beta$ parameter $\beta_{\rm edge} = 1.6045 \times 10^4$, an angular momentum parameter $\lambda_{\rm edge} = 6.396$, a specific energy parameter ${\varepsilon}_{\rm edge} = 1.7593 \times 10^{-3}$, and an accretion rate $\dot{m}_{\rm edge} = 0.07324$. The values of $\Phi_{\rm s}$ and $m$ used in this solution are shown in the plot.} 
	\label{fig_3}%
\end{figure}

As previously mentioned, the existence of an outer critical point in Fig.~\ref{fig_1}(a) is significant since this gives rise to the possibility of forming a shock wave in the AF. The complete accretion solution with a shock is shown in Figure~\ref{fig_3}, where the flow's Mach number ($M$) is plotted against the logarithmic radial distance $x$. $\beta_{\rm edge} = 1.6045 \times 10^4$, $\lambda_{\rm edge} = 6.396$, ${\varepsilon}_{\rm edge} = 1.7593\times10^{-3}$, and $\dot{m}_{\rm edge} = 0.07324$ are the values of the various flow variables at $x_{\rm edge} = 2000$, which is considered the outer edge of the AD. In addition, the wind parameter is considered with $m = 0.06$, and the SC parameter is set at $\Phi_{\rm s} = 1.5 \times 10^{-3}$. When these conditions exist at the outer edge, the subsonic AF approaches the BH horizon and crosses the outer critical point, shown by `O', to become supersonic. The outer critical point is located at $x_{\rm out}$ = 388.07155. A centrifugal force repels the supersonic flow as it gets closer to the BH horizon. Shock is produced in the vicinity of the BH as a result of the gravitational attraction and centrifugal barrier competing and the centrifugal barrier dominating over gravitational attraction \citep{Giri2015}. As a result of the shock transition, the PSC becomes hot, thick, and puffed up \citep{Chak1996b,2005A&AM,Aktar2015,Giri2015}. In this case, the shock is located at $x_s$ = 35.22 (as shown by the vertically downward arrow), and supersonic flow transitions into a subsonic flow via the shock transition. Shock waves are prevalent in astrophysical flows, and in a shock wave, the flow variables undergo a sharp and drastic jump. Certain shock conditions need to be satisfied to form a stable shock solution. In the description of the shock conditions, quantities with subscript `+' denote the PS parameters, and those with subscript `-' indicate the pre-shock (PRS) parameters. Following \citet{1959Landau,Sarkar2016,Sarkar2018,Sarkar2020RAA}, across the shock front, (a) the mass accretion rate has to be conserved (${\dot{M_+}=\dot{M_-}}$), (b) there has to be pressure balance across the shock (${W_++\Sigma _+u_+^2=W_-+\Sigma _-u_-^2}$), (c) the specific energy has to be conserved across the shock front (${\small{{\varepsilon}_+ = \varepsilon_-}}$), and also (d) the magnetic flux advection rate is conserved across the shock front (${\dot{\Phi}_+ = \dot{\Phi}_-}$). Following \citet{2014MNRASSam,Das-Sarkar2018,Sarkar2020RAA}, the flow's specific energy is calculated as 
$$
{\varepsilon}(x) = u^2/2+ c_s^2/(\gamma-1)+{\lambda^2}/{2x^2} +\Psi_{\rm BH} (x)+ {<B_\phi ^2>}/{(4\pi \rho)}.
\eqno(22)
$$ 
Further following \citet{Chakrabarti1996a,Gu2004ChPhL}, we define the flow's specific entropy as $s(x)$ $\propto$ ${c_s^{2n}}/{\rho}$ $\propto$ $u x c_s^{2n+1}\sqrt{\frac{\displaystyle x}{\displaystyle \gamma F}}$. The flow's kinetic energy transforms into enthalpy, and the ram pressure of the flow essentially changes to thermal pressure during a shock \citep{Frank2002,2007CC}. The laws of thermodynamics favour a discontinuous shock jump in the global transonic accretion solutions whenever it is dynamically possible since after the shock transition, the flow's entropy rises \citep{2007Pringle,2007CC,Kato2020}. Thus, the supersonic flow does not follow the black dot-dashed curve when shock conditions are satisfied. After passing past the inner critical point located at $x_{\rm in}$ = 3.3575 (shown by `I'), the subsonic PS flow once again becomes supersonic and enters the BH horizon.

\subsection{Shock dynamics as a function of $\Phi_{\rm s}$.}

\begin{figure}
	\centering 
	\includegraphics[height=8.0 cm,width=0.52\textwidth]{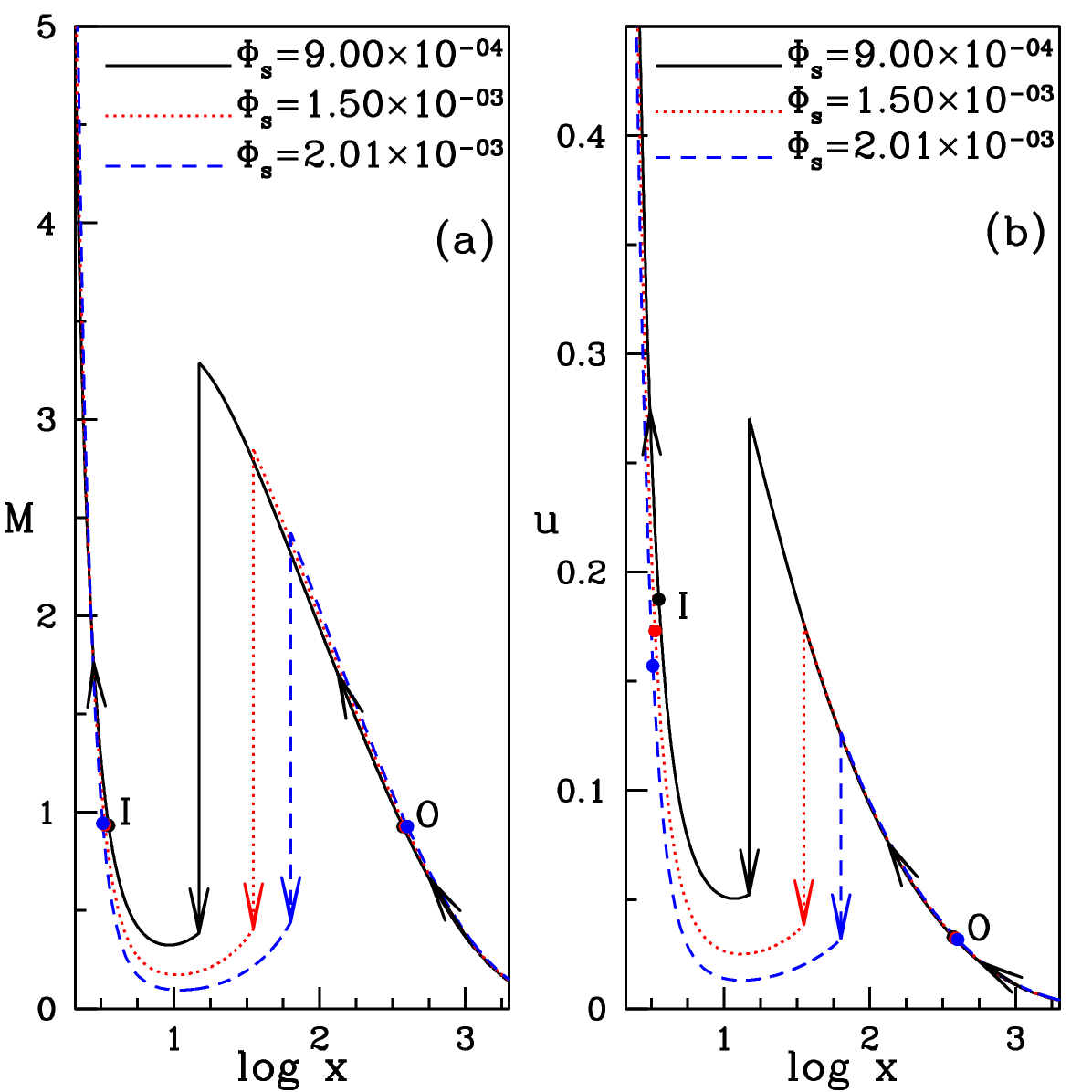}	
	\caption{(a) Mach number variation as a function of logarithmic radial distance. With energy ${\varepsilon}_{\rm edge}= 1.7593\times10^{-3}$, angular momentum $\lambda_{\rm edge} = 6.396$, $\beta_{\rm edge} = 1.6045 \times 10^4$, and $\dot{m}_{\rm edge} = 0.07324$ with fixed $m = 0.06$, the flow is injected from the outer edge $x_{\rm edge} = 2000$. The findings obtained for $\Phi_{\rm s}$ = $9.00 \times 10^{-4}$, $1.50 \times 10^{-3}$, and $2.01 \times 10^{-3}$ are shown as solid, dashed, and dot curves, respectively. The shock transitions at $x_s = 14.88$ (solid), $x_s$ = 35.22 (dotted), and $x_s$ = 63.46 (dashed) are indicated by vertical arrows. (b) The global shocked accretion solutions displayed in panel (a) correspond to the radial velocity ($u$) as a function of radial distance. The text has further information.} 
	\label{fig_4}%
\end{figure}

Figure~\ref{fig_4} shows the variation of shock location as a function of the SC parameter $\Phi_{\rm s}$ for the fixed parameters at the outer edge $x_{\rm edge} = 2000$, which are $\beta_{\rm edge} = 1.6045 \times 10^4$, $\lambda_{\rm edge} = 6.396$, ${\varepsilon}_{\rm edge} = 1.7593\times10^{-3}$, and $\dot{m}_{\rm edge} = 0.07324$, and for a fixed wind parameter $m = 0.06$. The findings for $\Phi_{\rm s}$ = $9.00 \times 10^{-4}$, $1.50 \times 10^{-3}$, and $2.01 \times 10^{-3}$ are shown in Fig.~\ref{fig_4}(a) as solid, dotted, and dashed curves, respectively, where the plot of the Mach number $M$ versus the logarithmic radial distance $x$ has been presented. The shock front moves away from the BH horizon as $\Phi_{\rm s}$ increases, as can be seen in the left panel, Fig.~\ref{fig_4}(a). It was reported by \citet{Faghei2012} that adding the SC parameter $\Phi_{\rm s}$ reduces the efficiency of $\lambda$ transmission. Therefore, the centrifugal barrier becomes more significant, and the shock front moves away from the BH horizon when $\Phi_{\rm s}$ is larger. In the right panel, Fig.~\ref{fig_4}(b), we have presented the variation of the radial flow velocity $u$ as a function of logarithmic radial distance $x$, corresponding to the shock solutions displayed in Fig.~\ref{fig_4}(a). It is seen that when $\Phi_{\rm s}$ increases, there is a drop in the PS radial flow velocity $u$ and a slowdown in the AF.

\subsection{Variation of various flow properties with respect to $\Phi_{\rm s}$, in accordance with Figure~\ref{fig_4}(a).}

\begin{figure}
	\centering 
	\includegraphics[height=9.0 cm,width=0.52\textwidth]{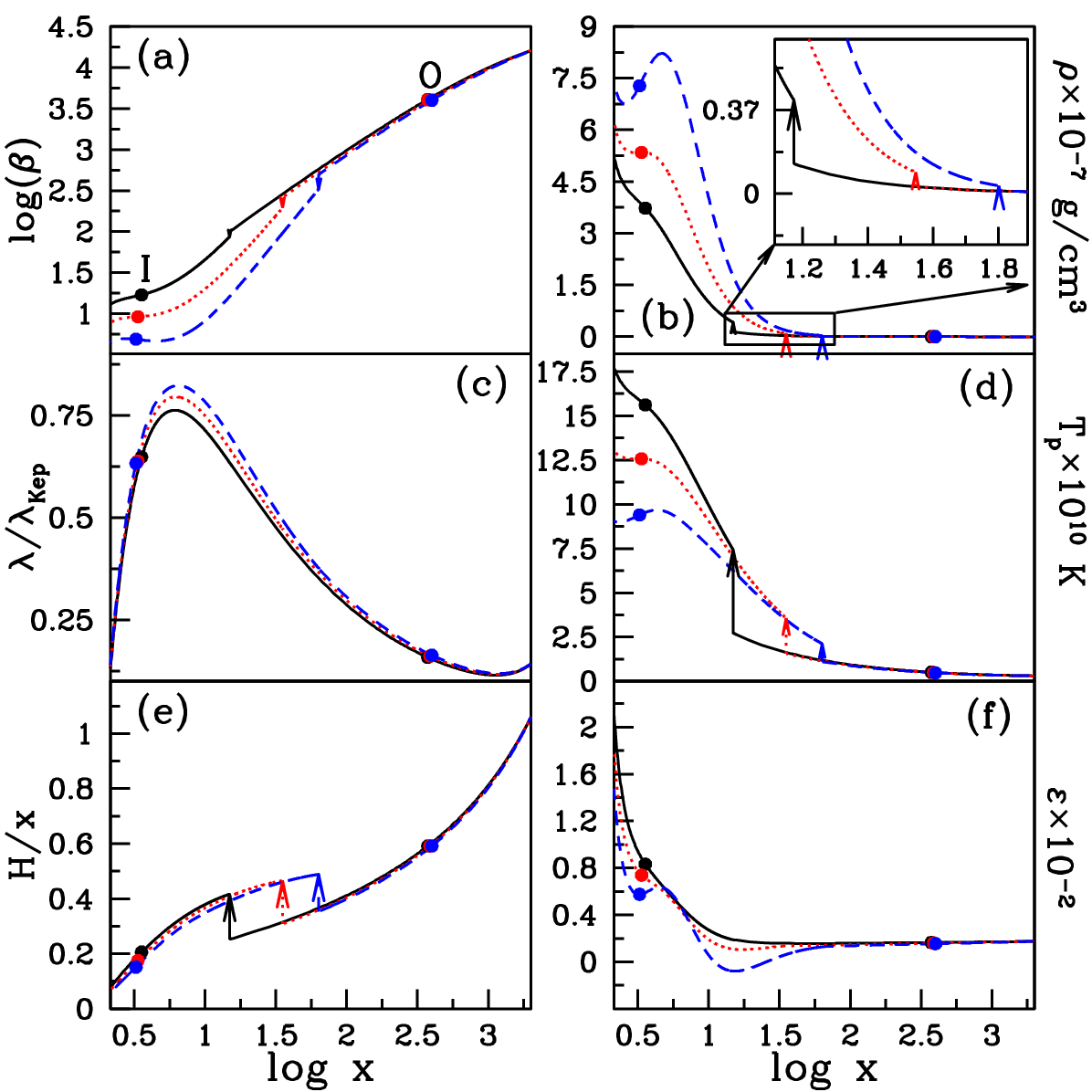}	
	\caption{Variation of the following as a function of logarithmic radial coordinate ($x$): (a) logarithmic ratio of gas pressure to magnetic pressure ($\beta$); (b) {\color{black}{normalized}} density ($\rho$) in g/cm$^3$; (c) {\color{black}{normalized}} angular momentum ($\lambda$); (d) {\color{black}{normalized}} proton temperature ($T_{\rm p}$); (e) disk aspect ratio ($H/x$); and (f) specific energy ($\varepsilon$). The solid, dotted, and dashed curves in Fig.~\ref{fig_4} reflect the accretion solutions. The critical points are shown as filled circles, with the inner critical point (I) being the closest circle and the outermost circle being the outer critical point (O). The vertical arrows indicate the shock position. The text contains details.}
	\label{fig_5}%
\end{figure}
\begin{figure}
	\centering 
	\includegraphics[height=9.0 cm,width=0.52\textwidth]{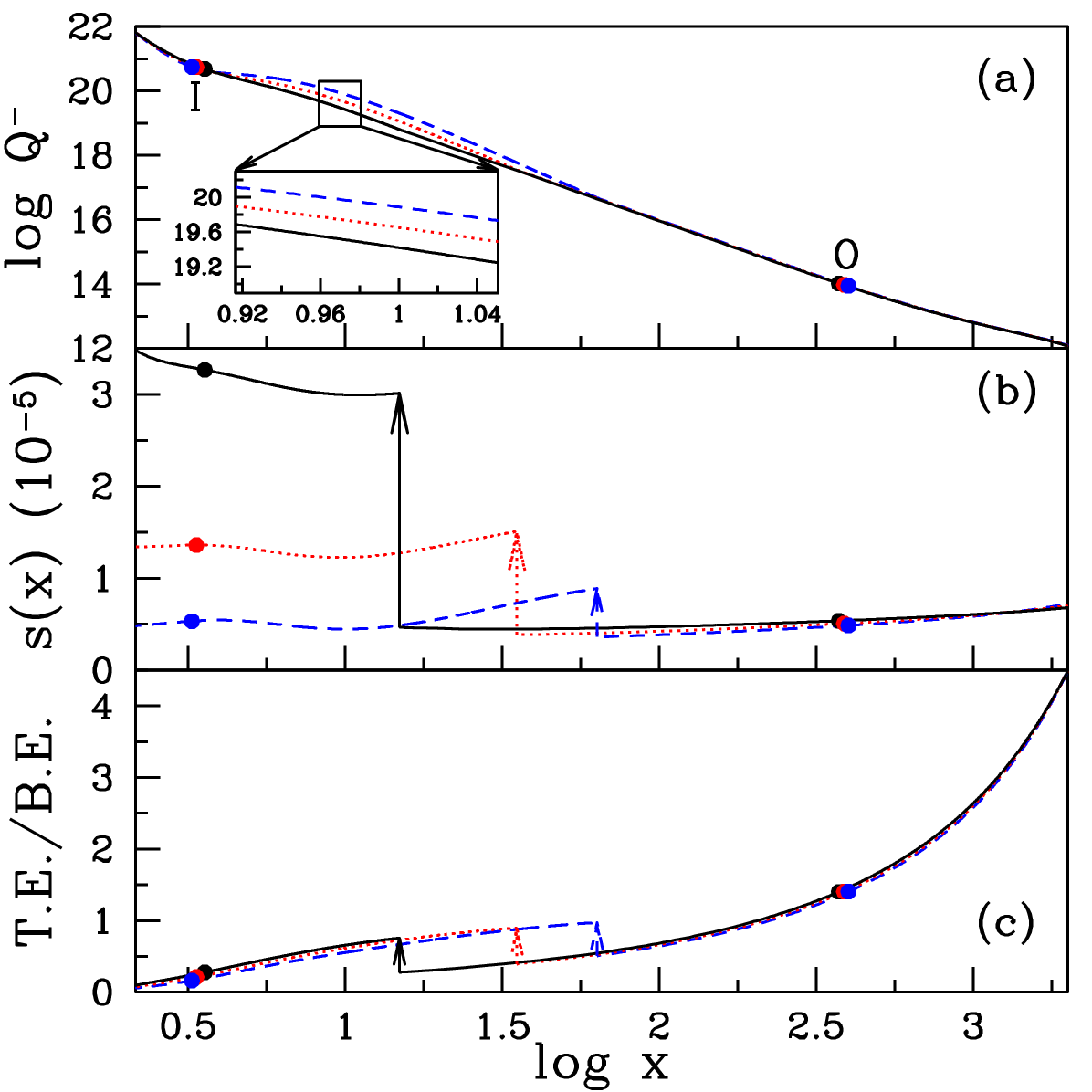}	
	\caption{{\color{black}{Variation of the following as a function of logarithmic radial coordinate ($x$): (a) logarithmic cooling rate ($Q^{-}$) in g/s$^3$; (b) entropy ($s(x)$); (c) ratio of thermal energy to binding energy (T.E./B.E.). The solid, dotted, and dashed curves in Fig.~\ref{fig_4} reflect the accretion solutions. The critical points are shown as filled circles, with the inner critical point (I) being the closest circle and the outermost circle being the outer critical point (O). The vertical arrows indicate the shock position. The text contains details.}}} 
	\label{fig_5_addl}%
\end{figure}

In Figure~\ref{fig_5}, we show the radial disk structure for the shock solutions that are presented in Fig.~\ref{fig_4}(a). In this case, we use 10 M$_\odot$ as the typical reference mass for the BH. The flow variable's radial variation is shown in each panel here. We plot the results using the same line styles as in Fig.~\ref{fig_4}(a). The filled dots show the critical points' locations. The variation of the logarithmic plasma$-\beta$ of the flow as a function of the logarithmic distance from the BH is shown in panel (a). Vertical arrows are used to depict the shock jumps. The fact that the plasma$-\beta$ drops in all three cases after the shock jump suggests that the PS flow is more magnetically dominated than the PRS flow. The density {\color{black}{(relative to $10^{-7}$ g/cm$^3$)}} variation of the AF as a function of radial distance is shown in panel (b). In every case, we discover that its density increases as the flow gets closer to the horizon.
Moreover, the density of the flow increases across the shock transition. An increase in $\Phi_{\rm s}$ results in a greater PS flow density since there is a slowdown in the AF with the increase of $\Phi_{\rm s}$ as reported in Fig.~\ref{fig_4}(b). The AF's specific angular momentum $\lambda$ {\color{black}{(relative to $\lambda_{\rm Kep}$)}} is then shown as a function of logarithmic radial distance in panel (c). The outward transport of $\lambda$ experiences a drop in efficiency as $\Phi_{\rm s}$ increases, which agrees with the findings of \citet{Faghei2012}. Panel (d) shows the radial variation in the proton temperature T$_p$ {\color{black}{(relative to $10^{10}$ K)}}, which shows a similar pattern to the density profile shown in panel (b). Nevertheless, for a larger value of $\Phi_{\rm s}$, T$_p$ in the PS region is lower. The disk aspect ratio's radial variation is displayed in panel (e). In every case, the aspect ratio is less than unity, except for the disk's outer edge. The PS region has substantial increases in disk aspect ratio when compared to the PRS region due to its high temperature and density. We depict the radial variation of the AF's specific energy in panel (f).

{\color{black}{We continue to discuss the radial disk structure for the shock solutions presented in Fig.~\ref{fig_4}(a), as shown in Fig.~\ref{fig_5_addl} for a 10 M$_\odot$ BH. Fig.~\ref{fig_5_addl} displays the variation of the following key quantities as a function of the logarithmic radial coordinate ($x$). Fig.~\ref{fig_5_addl}(a) illustrates the radial variation of logarithmic cooling rate ($Q^{-}$) in g/s$^3$, which increases steadily as the AF approaches the BH horizon. Notably, cooling rates ($Q^{-}$) remain continuous across the shock. Notably, our results indicate that, within the specified radial range ($6.0 \lesssim x \lesssim 60$), $Q^{-}$ is sensitively dependent on $\Phi_{\rm s}$, exhibiting an increase with rising $\Phi_{\rm s}$ values. In Fig.~\ref{fig_5_addl}(b), we present the entropy profile $s(x)$, demonstrating how heating and cooling processes influence flow properties. We observe an upward jump in entropy across the shock transition, indicating that shock transitions in the AF are thermodynamically favored. Moreover, as $\Phi_{\rm s}$ decreases, bringing the shock closer to the BH horizon, the entropy in the PS region is found to increase. Further, Fig.~\ref{fig_5_addl}(c) displays the ratio of thermal-to-binding energy (T.E./B.E.), defined as T.E./B.E. = $\frac{\displaystyle c_s^2/(\gamma - 1)}{\displaystyle |\Psi_{\rm BH}(x)|}$. Our results show that thermal energy dominates or is comparable to gravitational binding energy for a significant portion of the disk, suggesting the possibility of TDWs originating from the AF. Furthermore, for larger values of $\Phi_{\rm s}$, when the shock is located at a larger distance from the BH horizon, we find that the T.E./B.E. ratio exhibits an upward jump at an earlier point. Critical points are marked as filled circles, with the inner critical point (I) closest and the outermost circle representing the outer critical point (O). Vertical arrows indicate the shock position.}}

\begin{figure}
	\centering 
	\includegraphics[height=9.0 cm,width=0.52\textwidth]{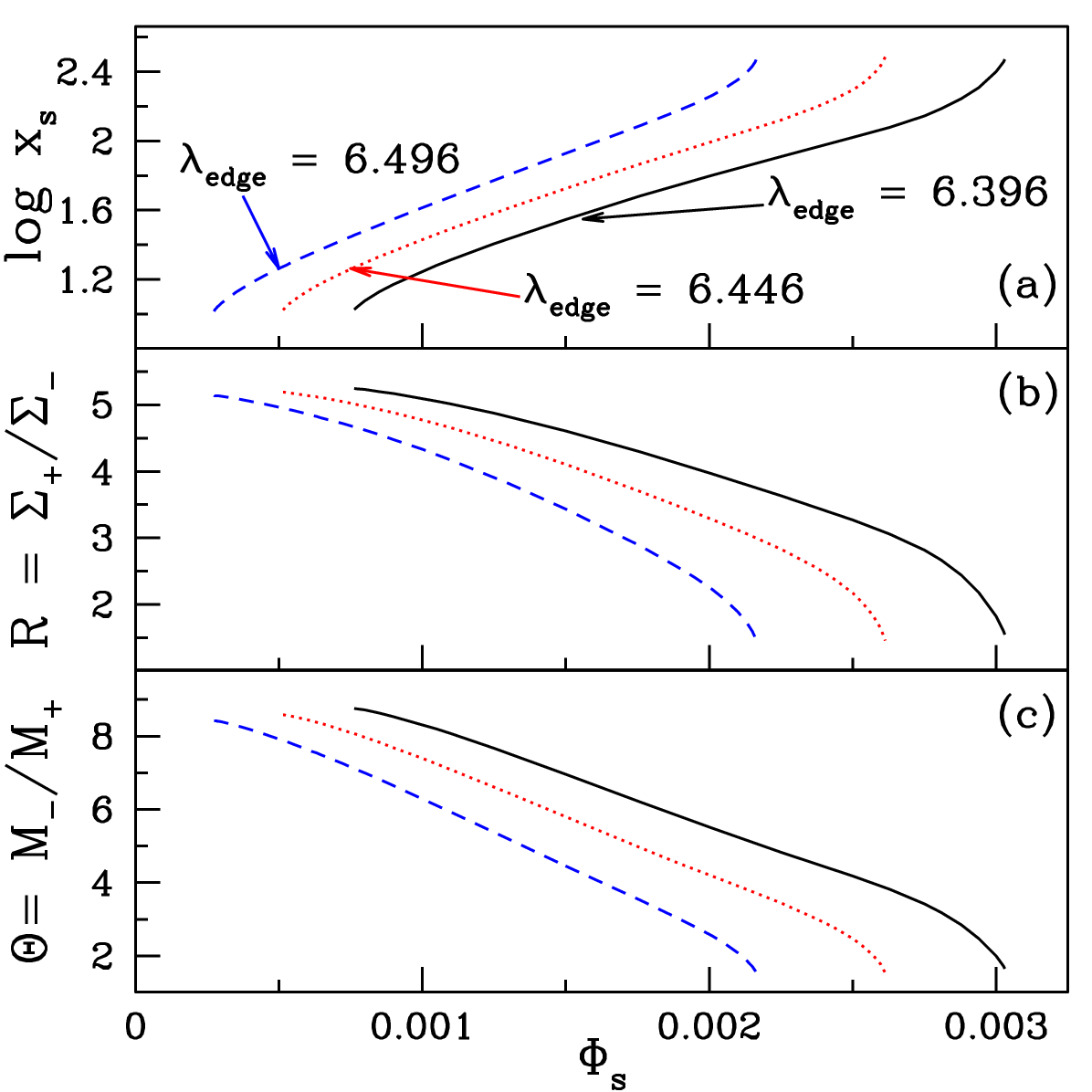}	
	\caption{For flows injected from $x_{\rm edge} = 2000$ with $\beta_{\rm edge} = 1.6045 \times 10^4$, ${\varepsilon}_{\rm edge} = 1.7593\times10^{-3}$, and $\dot{m}_{\rm edge} = 0.07324$, the variation as a function of $\Phi_{\rm s}$ of: (a) logarithmic shock location $x_s$, (b) compression ratio $R$, and (c) shock strength $\Theta$. The value of $m = 0.06$ is chosen as the mass-loss parameter. The curves that are solid, dotted, and dashed, respectively, represent the findings for $\lambda_{\rm edge}$ = 6.396, 6.446, and 6.496. The text contains details.}
	\label{fig_6}%
\end{figure}

\subsection{Variation of shock properties as a function of $\Phi_{\rm s}$}

The different shock properties for flows injected from a fixed outer edge are shown in Figure~\ref{fig_6} as a function of the SC parameter ($\Phi_{\rm s}$), where $x_{\rm edge} = 2000$, $\beta_{\rm edge} = 1.6045 \times 10^4$, specific energy ${\varepsilon}_{\rm edge} = 1.7593\times10^{-3}$, and $\dot{m}_{\rm edge} = 0.07324$. The mass-loss parameter is chosen as $m = 0.06$. The variation of the shock location for three distinct values of the angular momentum ($\lambda_{\rm edge}$) at $x_{\rm edge}$ is shown in the top panel (Fig.~\ref{fig_6}a). The solid, dotted, and dashed curves show the flows injected with angular momentum $\lambda_{\rm edge}= 6.396$, $6.446$, and $6.496$, respectively. The figure demonstrates that shock-induced global accretion solutions span an extensive $\Phi_{\rm s}$ range. The shock position moves farther from the BH horizon for a particular value of $\lambda_{\rm edge}$ as the SC parameter ($\Phi_{\rm s}$) increases. As previously mentioned in Fig.~\ref{fig_5}(c), an increase in $\Phi_{\rm s}$ results in a decrease in the outward transmission of $\lambda$ efficiency and an increase in centrifugal repulsion inside the flow. As a result, the shock front recedes from the horizon as $\Phi_{\rm s}$ increases. Standing shocks cannot form when the SC parameter crosses it's low ($\Phi_{\rm s}^{\rm cri,l}$) as well as high ($\Phi_{\rm s}^{\rm cri,h}$) critical values, since then the shock criteria are not fulfilled illustrating how the likelihood of a shock formation decreases beyond the critical limits of $\Phi_{\rm s}$. We should note that $\Phi_{\rm s}^{\rm cri}$ has no global value. Rather, AF parameters have a major role in this. Remarkably, in the case when $\Phi_{\rm s} < \Phi_{\rm s}^{\rm cri,l}$, there is no shock in the flow as well. Therefore, steady shock may occur in the AF throughout the range $\Phi_{\rm s}^{\rm cri,l} < \Phi_{\rm s} < \Phi_{\rm s}^{\rm cri,h}$. Moreover, an increase in $\lambda_{\rm edge}$ causes the shock front to recede from the horizon for a given $\Phi_{\rm s}$, which occurs due to the centrifugal barrier strengthening as $\lambda_{\rm edge}$ increases, demonstrating the centrifugally driven nature of the shocks.

Shock compression increases the density and temperature of the PS region, as has previously been mentioned. Furthermore, the flow's density and temperature distribution directly affect an AD's emitted spectrum. The density and temperature enhancement over the shock transition should thus be measured. To do this, we first calculate the compression ratio, which is shown as a function of $\Phi_{\rm s}$ in Fig.~\ref{fig_6}(b) and is defined as the ratio of the vertically averaged PS density to the PRS density ($R=\Sigma_{+}/\Sigma_{-}$). The flow parameters are selected as shown in Fig.~\ref{fig_6}(a). The value of $R$ decreases monotonically as $\Phi_{\rm s}$ increases for fixed values of $\lambda_{\rm edge}$. The decrease in the value of $R$ occurs because the rise in $\Phi_{\rm s}$ pushes the shock front farther from the BH horizon, reducing density compression and, as a result, lowering the compression ratio. Conversely, for constant $\Phi_{\rm s}$, increasing $\lambda_{\rm edge}$ reduces PS compression as the shock recedes because of the stronger centrifugal barrier. We see a low as well as high cut-off in $R$ for all the situations, as shock exists only for $\Phi_{\rm s}^{\rm cri,l} < \Phi_{\rm s} < \Phi_{\rm s}^{\rm cri,h}$. Next, we determine the shock strength ($\Theta$), which is simply a measurement of the temperature jump across the shock and is defined as the ratio of the PRS Mach number ($M_{-}$) to the PS Mach number ($M_{+}$) \citep{2007MNRASDas}. Fig.~\ref{fig_6}(c) displays the $\Theta$ variation as a function of $\Phi_{\rm s}$ for the same set of flow parameters as Fig.~\ref{fig_6}(a). Our findings reveal that $\Theta$ responds similarly to $R$ as seen in Fig.~\ref{fig_6}(b) when $\Phi_{\rm s}$ increases.

\subsection{Shock dynamics with respect to mass-loss parameter $m$}

\begin{figure}
	\centering 
	\includegraphics[height=8.0 cm,width=0.52\textwidth]{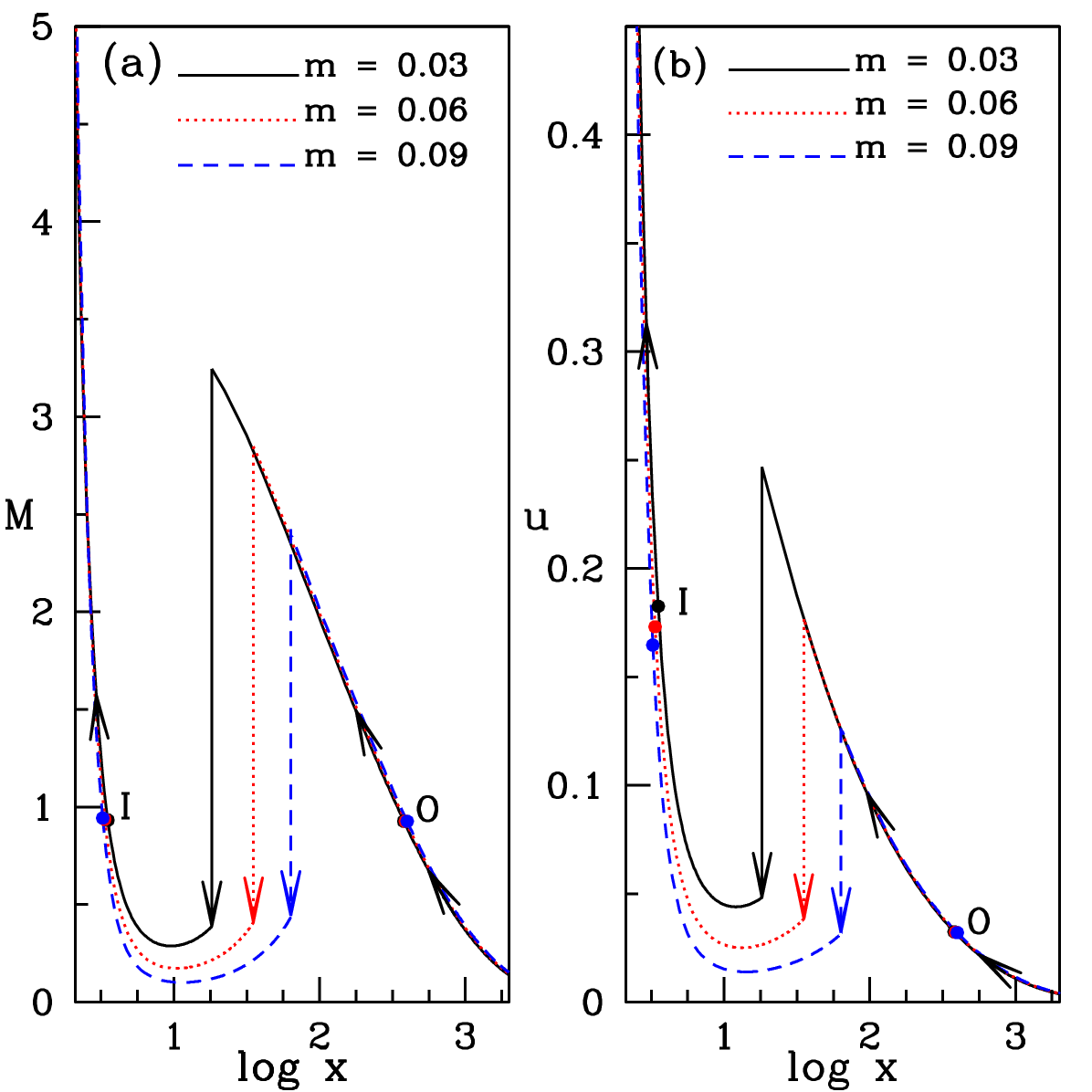}	
	\caption{(a) Mach number variation as a function of logarithmic radial distance. With energy ${\varepsilon}_{\rm edge} = 1.7593\times10^{-3}$, angular momentum $\lambda_{\rm edge} = 6.396$, $\beta_{\rm edge} = 1.6045 \times 10^4$, and $\dot{m}_{\rm edge} = 0.07324$, all fixed at $\Phi_{\rm s} = 1.50 \times 10^{-3}$, the flow is injected from the outer edge $x_{\rm edge} = 2000$. The findings obtained for $m$ = $0.03$, $0.06$, and $0.09$, respectively, are shown as solid, dotted, and dashed curves. The shock transitions at $x_s = 18.21$ (solid), $x_s$ = 35.22 (dotted), and $x_s$ = 64.55 (dashed) are indicated by vertical arrows. (b) The global shocked accretion solutions displayed in panel (a) correspond to the radial velocity ($u$) as a function of radial distance. The text has further information.} 
	\label{fig_7}%
\end{figure}

The variation of shock location as a function of the mass-loss parameter $m$ has been studied in Figure~\ref{fig_7} for the fixed parameters at the outer edge $x_{\rm edge} = 2000$, which are given by $\beta_{\rm edge} = 1.6045 \times 10^4$, $\lambda_{\rm edge} = 6.396$, ${\varepsilon}_{\rm edge} = 1.7593\times10^{-3}$, and $\dot{m}_{\rm edge} = 0.07324$, and for fixed SC parameter $\Phi_{\rm s} = 1.50 \times 10^{-3}$. The findings corresponding to $m$ = $0.03$, $0.06$, and $0.09$ are represented by solid, dotted, and dashed curves in Fig.~\ref{fig_7}(a), where the plot of the Mach number $M$ versus the logarithmic radial distance $x$ has been presented. We observe that the shock front moves away from the BH horizon as $m$ increases, as is evident from the left panel, Fig.~\ref{fig_7}(a). In the formalism of the present paper, we have assumed that the outflow originates from the AF without extracting $\lambda$. Thus, the mass loss will deposit $\lambda$ in the AF. So, the flow's $\lambda$ increases with the wind parameter $m$ increase. As the $m$ parameter increases, the centrifugal repulsion grows stronger, causing the shock location to move further away from the EH. We have illustrated the relationship between the radial flow velocity ($u$) and the logarithmic radial distance ($x$) in the right-hand panel, Fig.~\ref{fig_7}(b), which corresponds to the shock solutions shown in Fig.~\ref{fig_7}(a). We discover that the $u$ in the PS zone falls as $m$ increases, and the AF slows down.

\subsection{Variation of various flow properties in relation to the mass-loss parameter $m$, as shown in Fig.~\ref{fig_7}(a)}

\begin{figure}
	\centering 
	\includegraphics[height=9.0 cm,width=0.52\textwidth]{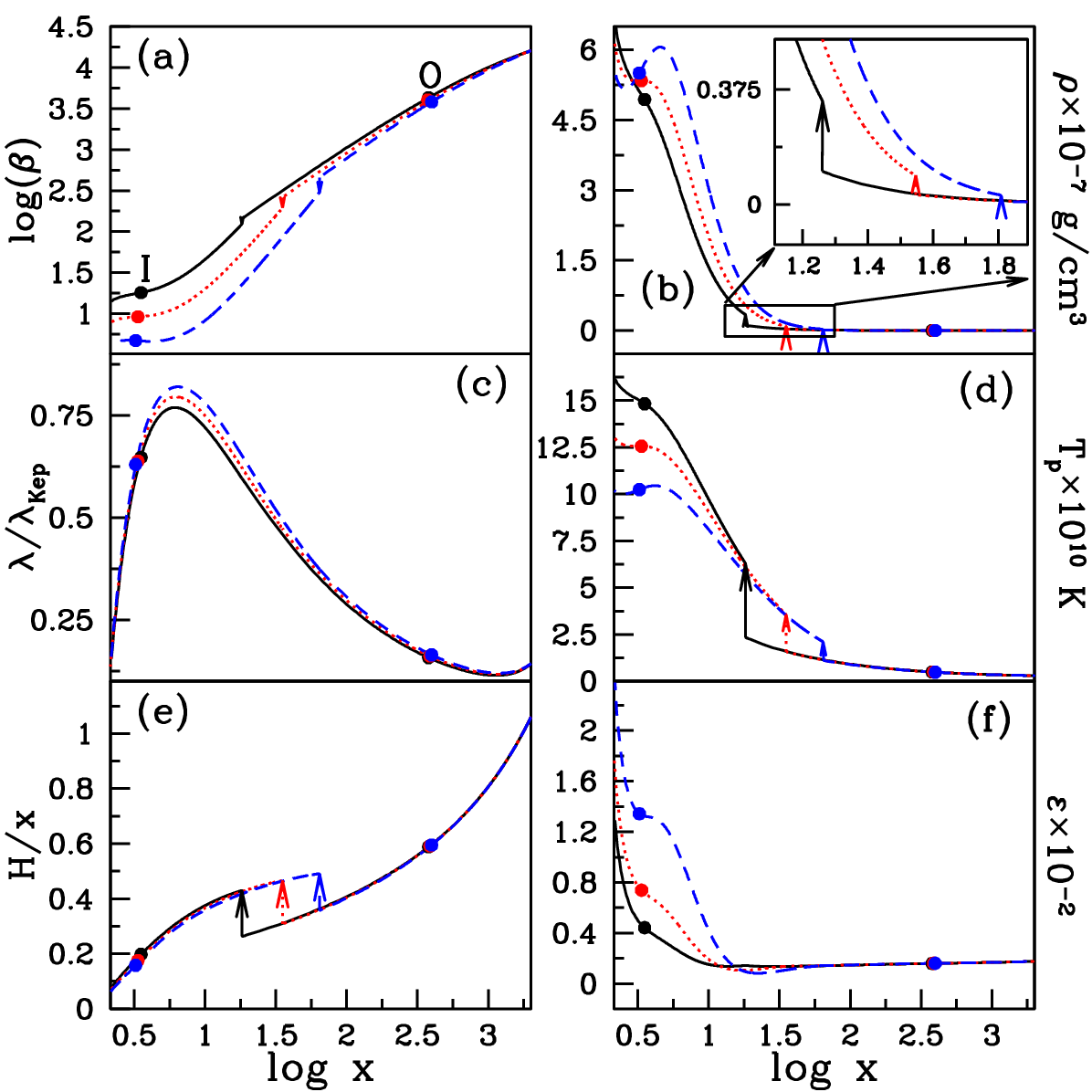}	
	\caption{Variation of the following as a function of logarithmic radial coordinate ($x$): (a) logarithmic ratio of gas pressure to magnetic pressure ($\beta$); (b) {\color{black}{normalized}} density ($\rho$) in g/cm$^3$; (c) {\color{black}{normalized}} angular momentum ($\lambda$); (d) {\color{black}{normalized}} proton temperature ($T_{\rm p}$); (e) disk aspect ratio ($H/x$); and (f) specific energy ($\varepsilon$). The solid, dotted, and dashed curves in Fig.~\ref{fig_7} reflect the accretion solutions. The critical points are shown as filled circles, with the inner critical point (I) being the closest circle and the outermost circle being the outer critical point (O). The vertical arrows indicate the shock position. The text contains details.} 
	\label{fig_8}%
\end{figure}
\begin{figure}
	\centering 
	\includegraphics[height=9.0 cm,width=0.52\textwidth]{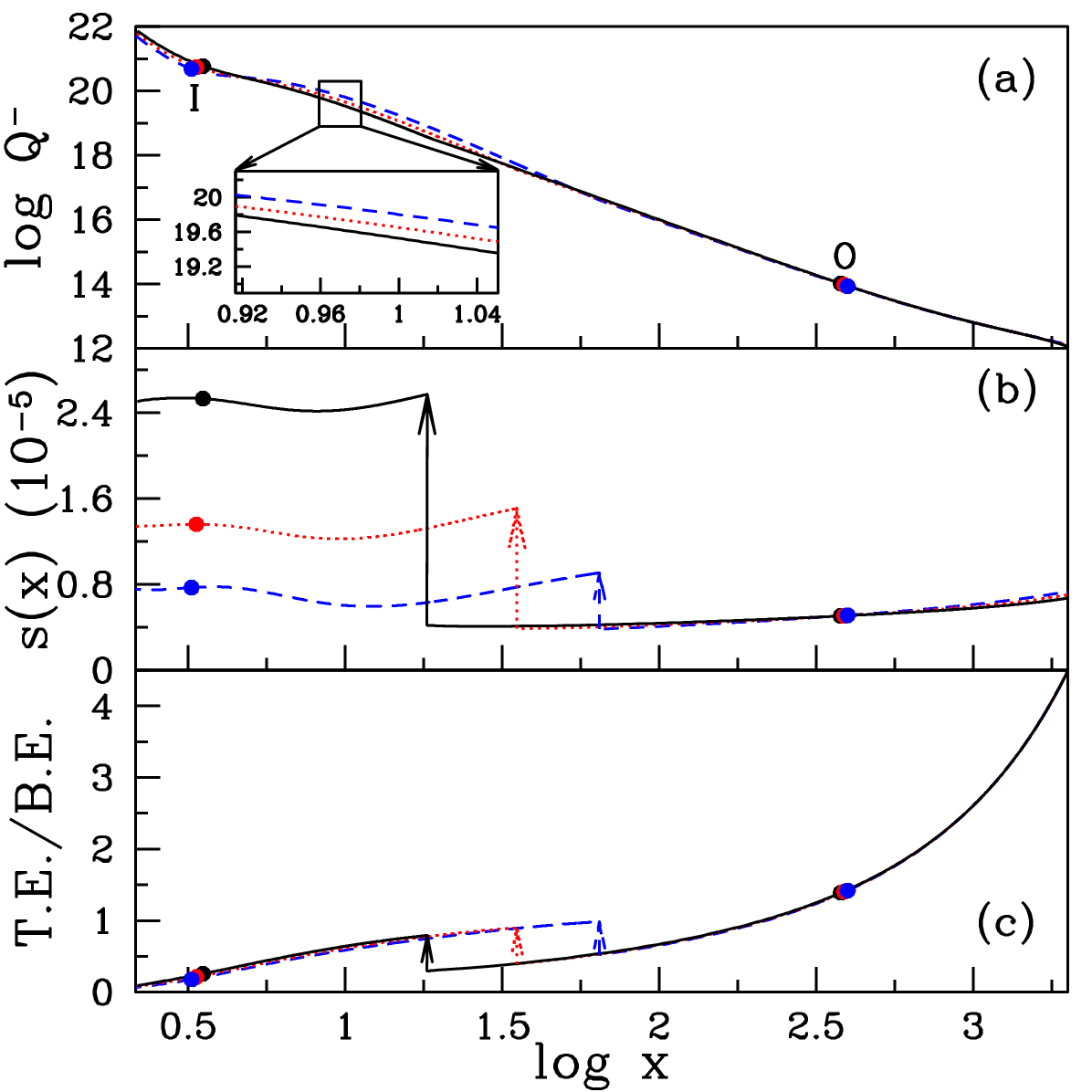}	
	\caption{{\color{black}{Variation of the following as a function of logarithmic radial coordinate ($x$): (a) logarithmic cooling rate ($Q^{-}$) in g/s$^3$; (b) entropy ($s(x)$); (c) ratio of thermal energy to binding energy (T.E./B.E.). The solid, dotted, and dashed curves in Fig.~\ref{fig_7} reflect the accretion solutions. The critical points are shown as filled circles, with the inner critical point (I) being the closest circle and the outermost circle being the outer critical point (O). The vertical arrows indicate the shock position. The text contains details.}}} 
	\label{fig_8_addl}%
\end{figure}

In Figure~\ref{fig_8}, we show the radial disk structure that corresponds to the shock solutions that are taken into account in Fig.~\ref{fig_7}(a). In this case, we use 10 M$_\odot$ as the typical reference mass for the BH. The radial variation of a flow variable is shown in each panel here. The same line styles as in Fig.~\ref{fig_7}(a) are used to illustrate the results. The filled dots show the critical points' locations. The variation of the logarithmic plasma$-\beta$ of the flow as a function of the logarithmic distance from the BH is shown in panel (a). Vertical arrows are used to depict the shock transitions. The fact that the plasma$-\beta$ decreases in all three cases after the shock jump suggests that the PS flow is more magnetically dominated than the PRS flow. As a function of radial distance, we display the density {\color{black}{(relative to $10^{-7}$ g/cm$^3$)}} variation of the AF in panel (b). In each case, the flow's density increases as it gets closer to the BH horizon.
Additionally, via the shock transition, the density is increased. A more significant value of $m$ corresponds to a higher PS flow density since there is a slowdown in the AF with the increase of $m$ as reported in Fig.~\ref{fig_7}(b). The variation of the AF's specific angular momentum $\lambda$ {\color{black}{(relative to $\lambda_{\rm Kep}$)}} as a function of radial distance is then shown in panel (c). The effectiveness of the outward transport of $\lambda$ diminishes as $m$ increases. Therefore, as $m$ increases, the shock front moves farther from the BH horizon due to a more substantial centrifugal barrier, as discussed in Fig.~\ref{fig_7}(a). Similar trends are displayed in the density profile presented in panel (b) and the proton temperature T$_p$ {\color{black}{(relative to $10^{10}$ K)}} profile in panel (d). For a more significant value of $m$, the PS flow's T$_p$ is, nevertheless, lower. We plot the disk aspect ratio's radial variation in panel (e). In all cases, the aspect ratio is less than unity, except the disk's outer edge. The high temperature and density in the PS region cause it to puff up, and when compared to the PRS region, the aspect ratio in the PS region increases significantly. We depict the radial variation of the AF's specific energy in panel (f).

{\color{black}{We examine the radial disk structure for shock solutions presented in Fig.~\ref{fig_7}(a), with detailed profiles shown in Fig.~\ref{fig_8_addl} for a 10 M$_\odot$ BH. Fig.~\ref{fig_8_addl} presents the radial dependence of key quantities versus logarithmic coordinate ($x$). Specifically, Fig.~\ref{fig_8_addl}(a) shows the steady increase in logarithmic cooling rate ($Q^{-}$) as the AF approaches the BH horizon, with continuous cooling rates across the shock. The entropy profile $s(x)$ in Fig.~\ref{fig_8_addl}(b) reveals the significant impact of heating and cooling processes on flow properties, characterized by an upward entropy jump across the shock transition. This indicates thermodynamically favorable shock transitions in the AF. Fig.~\ref{fig_8_addl}(c) displays the thermal-to-binding energy ratio (T.E./B.E.), defined in Section 3.4. Our analysis reveals that thermal energy is comparable to or dominates gravitational binding energy over a substantial portion of the disk, implying the potential for TDWs to originate from the AF. Critical points are denoted by filled circles, with inner (I) and outer (O) points marked. Vertical arrows indicate shock positions. Moreover, the curves in panels (a), (b), and (c) of Fig.~\ref{fig_8_addl} exhibit similar variational characteristics with increasing $m$ parameter, analogous to those observed in Fig.~\ref{fig_5_addl}, where the dependence on the $\Phi_{\rm s}$ parameter was investigated.}}

\subsection{Variation of shock properties as a function of mass-loss parameter $m$}

\begin{figure}
	\centering 
	\includegraphics[height=9.0 cm,width=0.52\textwidth]{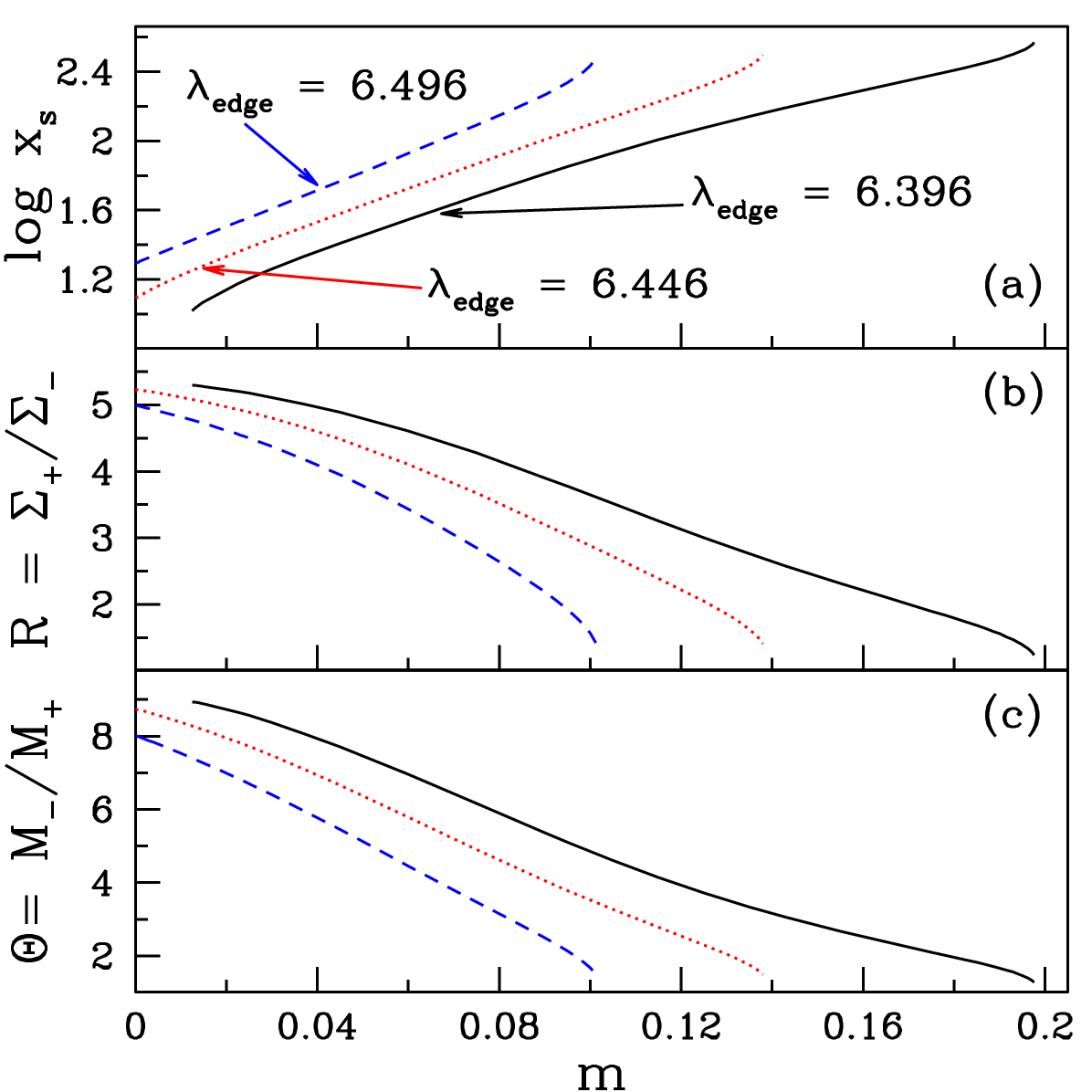}	
	\caption{Variation of (a) logarithmic shock location $x_s$, (b) compression ratio $R$, and (c) shock strength $\Theta$ for flows injected from $x_{\rm edge} = 2000$ with $\beta_{\rm edge} = 1.6045 \times 10^4$, ${\varepsilon}_{\rm edge} = 1.7593\times10^{-3}$, and $\dot{m}_{\rm edge} = 0.07324$; all as a function of mass-loss parameter $m$. The chosen value for $\Phi_{\rm s}$ is $1.5 \times 10^{-3}$. The curves that are solid, dotted, and dashed, respectively, represent the findings for $\lambda_{\rm edge}$ = 6.396, 6.446, and 6.496. The text contains details.} 
	\label{fig_9}%
\end{figure}

Then, for flows with the same outer boundary values — that is, $x_{\rm edge} = 2000$, ${\varepsilon}_{\rm edge} = 1.7593\times 10^{-3}$, $\beta_{\rm edge} = 1.6045 \times 10^4$, and $\dot{m}_{\rm edge} = 0.07324$ — we continue to investigate the shock properties in terms of wind parameter $m$ in Figure~\ref{fig_9}. $\Phi_{\rm s}$ is fixed at $1.5 \times 10^{-3}$. The solid, dotted, and dashed curves show the results that correspond to $\lambda_{\rm edge}= 6.396$, $6.446$, and $6.496$, respectively. In all situations with given $\lambda_{\rm edge}$, the shock front recedes away from the horizon with an increase of $m$. As previously mentioned in Fig.~\ref{fig_8}(c), when $m$ is increased, the effectiveness of the outward transport of $\lambda$ diminishes, and therefore, shock moves outward due to an increase in centrifugal repulsion. However, shock ceases to exist beyond a critical limit ($m^{\rm cri}$). Therefore, an infinite rise in $m$ is not achievable while maintaining the other flow parameters unaltered. Here, as in Fig.~\ref{fig_6}(b-c), we examine how the shock strength ($\Theta$) and compression ratio ($R$) vary with $m$. With regard to $m$, we discover that $R$ and $\Theta$ both exhibit an anti-correlated relationship.

\subsection{{\color{black}{Estimation of mean free path of electrons, Bondi radius, and disk height}}}

{\color{black}{According to \citet{1998ApJ_Quataert,2021Ap&SS_Ghoreyshi,Mitra2023}, in hot AFs, the mean free path of Coulomb interactions exceeds the system's characteristic size, rendering the ionic-electronic coupling too weak, and thus these flows behave as essentially collisionless systems. Moreover, the thermal conductivity enters a `saturated' regime when the electron mean free path ($l_{\rm mfp}$) approaches or surpasses the characteristic length scale of the temperature gradient given by $l_T$ = $T/|\nabla T|$ \citep{Cowie-McKee1977,1986RvMP1S,1986ApJB,Tanaka-Menou2006,2008MNRASA,Faghei2012,Rezgui2019ApJ,2023MNRASS}. If we approximate $\nabla T$ in 1D as $T/x$, then $l_T \sim x$ \citep{Tanaka-Menou2006}. We need to multiply $l_T$ by $r_g$ to make it dimensionful, where $r_g$ denotes the Schwarzschild radius. Considering a 10 $M_{\odot}$ BH, we find $r_g \simeq 1.5 \times 10^6$ cm.}}

{\color{black} In order to justify our approach to TC, we proceed to estimate the electron mean free path ($l_{\rm mfp}$) and contrast it with key physical length scales relevant to the problem, namely the Bondi radius ($r_B$) and disk thickness ($2H$).}

\subsubsection{{\color{black}{Estimation of mean free path of electrons, $l_{\rm mfp}$ (at $x_{\rm in}$):}}}

{\color{black}{Following \citet{Frank2002,2007Pringle,Courvoisier2013}, we calculate the mean free path of electrons ($l_{\rm mfp}$) at the inner sonic point ($x_{\rm in}$) for a representative case depicted in Fig.~\ref{fig_4}(a), corresponding to $\Phi_{\rm s} = 9.0 \times 10^{-4}$. To estimate this, we use the vertically integrated density $\Sigma = 2\rho I_n H$, where $I_n = (2^n n!)^2 /(2n + 1)!$ and $n = \frac{\displaystyle 1}{\displaystyle \gamma - 1}$ is the polytropic index \citep{Chakrabarti1989}. For a disk of thickness $2H$, the electron number density ($n$, in cm$^{-3}$) is calculated as $n = \frac{\displaystyle \Sigma}{\displaystyle 2H m_p} = \frac{\displaystyle \rho I_n}{\displaystyle m_p}$. Thereafter, $l_{\rm mfp}$ (in cm) is estimated using the relation $l_{\rm mfp} = \frac{\displaystyle 7 \times 10^{5}}{\displaystyle ln \Lambda} \frac{\displaystyle T^2}{\displaystyle n}$. For the representative case considered, where $x_{\rm in}$ = 3.56826, we have $\rho_{\rm in} = 10^{-6.43}$ g/cm$^3$, $m_p = 1.673 \times 10^{-24}$ g, and $\gamma$ = 4/3. Furthermore, the Coulomb logarithm, denoted as $ln \Lambda$, is a numerical factor typically on the order of 10. For calculating the $l_{\rm mfp}$ of electrons, $T$ in the expression of $l_{\rm mfp}$ is considered as the electron temperature, $T_e = (\sqrt{m_e/m_p})T_p$ \citep{CC2002}. The plasma temperature at $x_{\rm in}$ is assumed to be equivalent to the proton temperature ($T_{\rm p, in}$), which has the specified value of $T_{\rm p, in} = 10^{11.19}$ K. So, the electron temperature at $x_{\rm in}$ is $T_{\rm e, in} = (\sqrt{m_e/m_p})T_{\rm p, in} = 2.334\times 10^{-2} \times 10^{11.19}$ K = $3.615 \times 10^9$ K. So, at $x_{\rm in}$ = 3.56826 $(\simeq 5.35 \times 10^6$ cm), the mean free path of electrons is $l_{\rm mfp, in} = \frac{\displaystyle 7 \times 10^{5}}{\displaystyle ln \Lambda} \frac{\displaystyle T_{\rm e, in}^2 m_p}{\displaystyle \rho_{\rm in} I_n}$ = $\frac{\displaystyle 7 \times 10^{5}}{\displaystyle 10} \frac{\displaystyle (3.615 \times 10^9)^2 \times 1.673 \times 10^{-24}}{\displaystyle 10^{-6.43} \times 0.45714}$ cm = $9.0 \times 10^{6}$ cm. So, we find that $l_{\rm mfp, in} \sim x_{\rm in}r_g$. Thus, the assumption of saturated TC in our model formalism is justified.}}

\subsubsection{{\color{black}{Estimation of mean free path of electrons, $l_{\rm mfp}$ (at $x_{\rm out}$):}}}

{\color{black}{For the representative case considered in Fig.~\ref{fig_4}(a), corresponding to $\Phi_{\rm s} = 9.0 \times 10^{-4}$, where $x_{\rm out}$ = 373.58835, we have $\rho_{\rm out} = 10^{-10.05}$ g/cm$^3$, $m_p = 1.673 \times 10^{-24}$ g, and $\gamma$ = 4/3. The plasma temperature at $x_{\rm out}$ is assumed to be equivalent to the proton temperature ($T_{\rm p, out}$), which has the specified value of $T_{\rm p, out} = 10^{9.71}$ K. So, the electron temperature at $x_{\rm out}$ is $T_{\rm e, out} = (\sqrt{m_e/m_p})T_{\rm p, out} = 2.334\times 10^{-2} \times 10^{9.71}$ K = $1.197 \times 10^8$ K. So, at $x_{\rm out}$ = 373.58835 $(\simeq 5.60 \times 10^8$ cm), the mean free path of electrons is $l_{\rm mfp, out} = \frac{\displaystyle 7 \times 10^{5}}{\displaystyle ln \Lambda} \frac{\displaystyle T_{\rm p, out}^2 m_p}{\displaystyle \rho_{\rm out} I_n}$ = $\frac{\displaystyle 7 \times 10^{5}}{\displaystyle 10} \frac{\displaystyle (1.197 \times 10^8)^2 \times 1.673 \times 10^{-24}}{\displaystyle 10^{-10.05} \times 0.45714}$ cm = 4.12 $\times 10^{7}$ cm. This result also confirms that $l_{\rm mfp, out}$ is close to $x_{\rm out}r_g$, further validating our assumption of saturated TC.}}

\subsubsection{{\color{black}{Estimation of Bondi radius, $r_B$:}}}

{\color{black}{We estimate the Bondi radius ($r_B$) as $r_B = GM_{\rm BH}/c{_s}(\infty)^2$, with $M_{\rm BH}$ being the BH mass and $c{_s}(\infty)$ representing the sound speed of the ambient gas at infinite distance \citep{Frank2002,2011MNRAS_Bu,2015Russell}. Following \citet{2011MNRAS_Bu}, we consider $c{_s}(\infty)^2 = 10^{-6} c^2$, where $c$ is light's speed. Thus, $r_B = 10^6\: GM_{\rm BH}/c^2 = 10^6\: r_g \simeq 10^{12}$ cm. Thus, we find that $l_{\rm mfp} << r_B$.}}

\subsubsection{{\color{black}{Estimation of disk thickness, $2H$ (at $x_{\rm in}$):}}}

{\color{black}{We now proceed to calculate the disk thickness ($2H$) at $x_{\rm in}$ for representative case considered in Fig.~\ref{fig_4}(a), using the radial variation of the disk aspect ratio ($H/x$) shown in Fig.~\ref{fig_5}(e). At $x_{\rm in}$ = 3.56826 (corresponding to the representative case considered in Fig.~\ref{fig_4}(a)), the disk aspect ratio is $(H/x)_{\rm in} = 0.20587$. Thus, the disk thickness at the inner sonic point is $2H_{\rm in} = 2 \times 0.20587 \times x_{\rm in} \times r_g = 2.2 \times 10^6$ cm. So, we find that $l_{\rm mfp, in} \sim 2H_{\rm in}$.}}

\subsubsection{{\color{black}{Estimation of disk thickness, $2H$ (at $x_{\rm out}$):}}}

{\color{black}{Finally, we calculate the disk thickness ($2H$) at $x_{\rm out}$ for representative case considered in Fig.~\ref{fig_4}(a), using the radial variation of the disk aspect ratio ($H/x$) shown in Fig.~\ref{fig_5}(e). At $x_{\rm out}$ = 373.58835 (corresponding to the representative case considered in Fig.~\ref{fig_4}(a)), the disk aspect ratio is $(H/x)_{\rm out} = 0.59140$. Thus, the disk thickness at the outer sonic point is $2H_{\rm out} = 2 \times 0.59140 \times x_{\rm out} \times r_g = 6.6282 \times 10^8$ cm. So, we also find that $l_{\rm mfp, out}$ is close to $2H_{\rm out}$.}}

\subsection{{\color{black}{Explaining the high gas temperature at the outer boundary}}}

{\color{black}{According to \citet{Kumar2021ApJ}, the advective AF around a BH may contain either cold-mode or hot-mode gas inflows. \citet{Kumar2021ApJ} observed that the hot-mode gas is more effective than the cold-mode gas in transporting angular momentum around the outer accretion boundary. A smooth global solution (often referred to as ADAF) or two sonic point solutions may be obtained from the cold-mode gas, while multiple global (also known as shock solutions) or single sonic point solutions can be obtained from the hot-mode gas.}}

{\color{black}{It is well known that Roche-lobe overflow occurs in low-mass BH X-ray binaries, and the companion star's overflow mostly feeds the BH, whereas the companion star's winds primarily feed the BHs of high-mass BH X-ray binaries. Thus, according to \citet{Kumar2021ApJ}, near the accretion boundary, the temperature of the accreting gas may vary. Failed winds from the ADAF or Keplerian disks and stellar winds might be the main sources of hot accreting gas in X-ray binaries \citep{Kumar2021ApJ}. Moreover, within high-mass BH X-ray binaries, direct accretion of the massive companion's hot stellar wind onto the BH gives rise to an ADAF-like hot AF, contingent upon maintaining a gas supply rate beneath the critical threshold for ADAF formation \citep{2022iSci_Liu,2024MNRAS_Wang}.}}

{\color{black}{The ions essentially retain the heat dissipated due to viscosity in the AF \citep{2007ApJ_Sharma,Sarkar2020RAA}. The electron-ion bremsstrahlung effectively reduces the electron temperature due to electrons' smaller mass compared to ions \citep{1986rpa_book_Rybicki}. According to \citet{CC2002}, the electron temperature is expected to be $\sqrt{m_e/m_p} \simeq 2.33 \times 10^{-2}$ times lower than the ion temperature. When the ion (proton) temperature greatly exceeds the electron temperature, ions predominantly contribute to the overall pressure, owing to the electrons' relatively low pressure \citep{2007ApJ_Sharma,2021MNRAS_Mizuno}. Our work demonstrates the radial variation of proton temperature (shown in Fig.~\ref{fig_5}(d) and Fig.~\ref{fig_8}(d)), explaining the assumed high gas temperature ($\sim$ 10$^{9.5}$ K) at the outer boundary.}}

\begin{figure}
	\centering 
	\includegraphics[height=8.0 cm,width=0.52\textwidth]{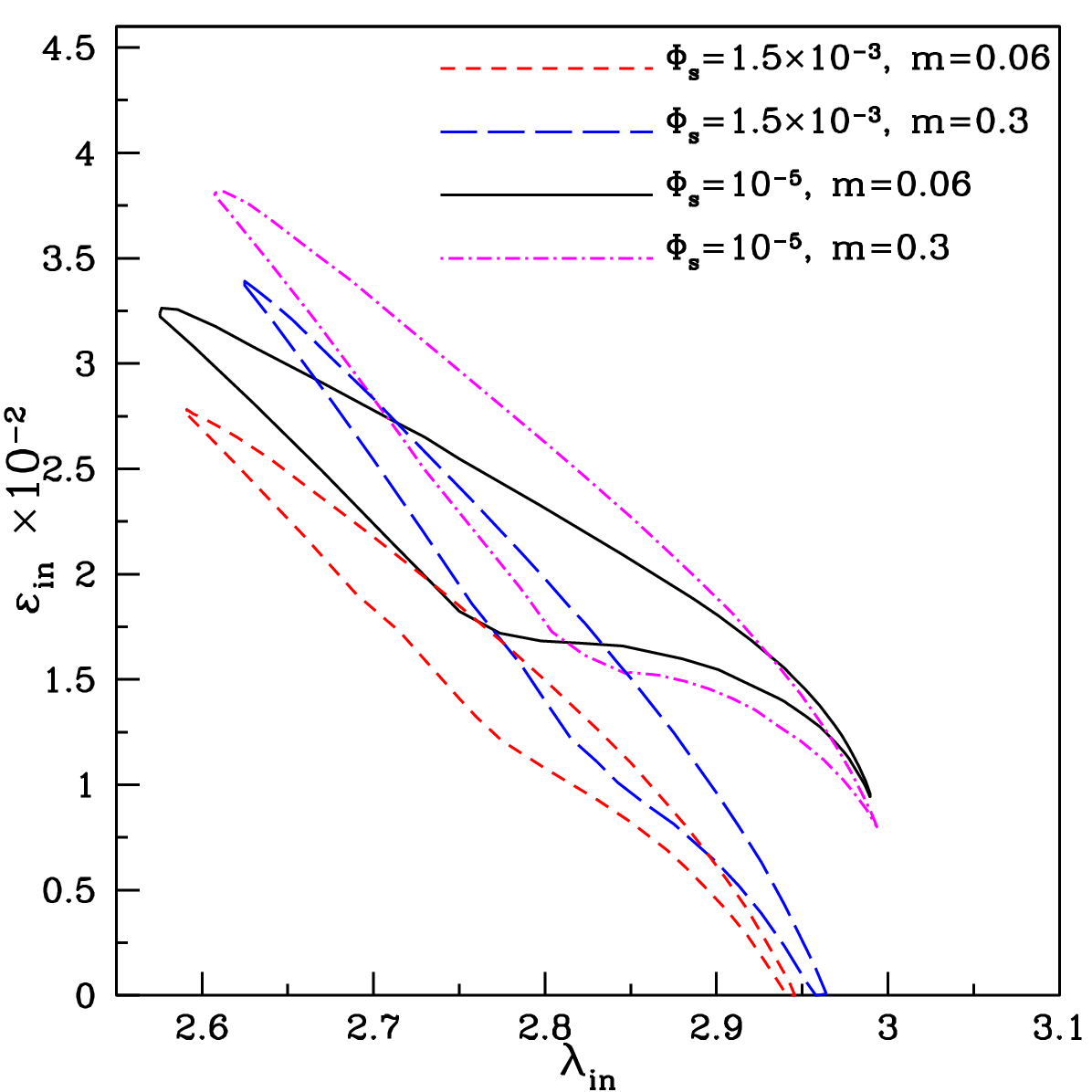}	
	\caption{As a function of the SC parameter ($\Phi_{\rm s}$) and wind parameter ($m$), the plot displays the modification in specific energy at the inner critical point ($\varepsilon_{\rm in}$) versus $\lambda$ at the inner critical point ($\lambda_{\rm in}$) in the parameter space for standing shocks. We fixed $\beta_{\rm in} = 10$ and ${\dot{m}}_{\rm in} = 0.05$ to generate this plot.} 
	\label{fig_10}%
\end{figure}
\begin{figure}
	\centering 
	\includegraphics[height=8.0 cm,width=0.52\textwidth]{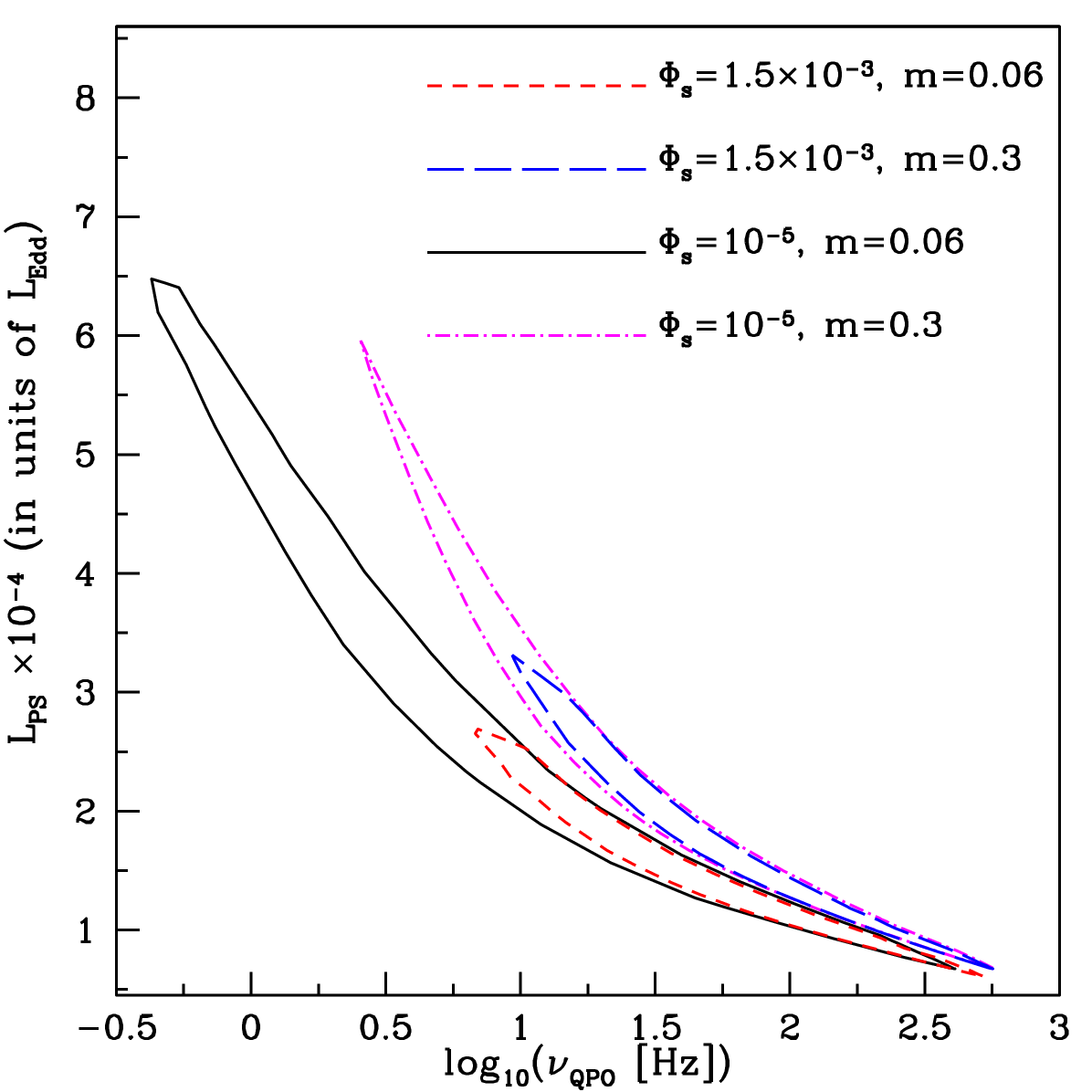}	
	\caption{This plot displays the parameter space of PS luminosity versus QPO frequency associated with the parameter space of $\varepsilon_{\rm in}$ vs $\lambda_{\rm in}$ presented in Fig.~\ref{fig_10}. Here, we have assumed a BH of mass $M_{\rm BH}$ = 10 $M_{\odot}$. See text for details.} 
	\label{fig_11}%
\end{figure}

\subsection{Parameter space study of the stationary shock solutions}

To comprehend the magnetized dissipative AF, in Fig.~\ref{fig_10}, we categorize the parameter space as a function of the $\Phi_{\rm s}$ and $m$ parameters in the $\varepsilon_{\rm in}-\lambda_{\rm in}$ plane, where $\varepsilon_{\rm in}$ indicates the specific energy of the flow at the inner critical point, $x_{\rm in}$. In doing so, we choose $\beta_{\rm in} = 10$ and ${\dot{m}}_{\rm in} = 0.05$ as representative values and freely varied the values of $x_{\rm in}$ and $\lambda_{\rm in}$. We demonstrate how the $\Phi_{\rm s}$ and $m$ parameters affect the standing shock formation process, which is represented by the $\varepsilon_{\rm in}-\lambda_{\rm in}$ parameter space. It is anticipated that increasing the $\Phi_{\rm s}$ and $m$ parameters will modify the shock parameter space. By varying the TC and wind parameters, we determine the altered regions of the parameter space that sustain stationary shock. In the Fig.~\ref{fig_10}, the $\Phi_{\rm s}$ and $m$ parameters are shown. When the SC parameter ($\Phi_{\rm s}$) increases for a fixed wind parameter $m$, the shock parameter space shifts to lower energies. Since the increase of $\Phi_{\rm s}$ causes the transfer of heat from the inner region to the outer region of the AF \citep{Ghasemnezhad2018}, this can explain the shift of the shock parameter space to lower energy when $\Phi_{\rm s}$ is increased. For a fixed $\Phi_{\rm s}$ parameter, when the wind parameter $m$ is increased, we find that the shock parameter space is shifted to higher specific energy $\varepsilon_{\rm in}$. As evident from Fig.~\ref{fig_8}(d), the outflow affects the AF's temperature profile and the plasma$-\beta$ profile (see Fig.~\ref{fig_8}(a)), which can explain the shift of the parameter space to higher energies for higher $m$ values. So, the $\Phi_{\rm s}$ and $m$ parameters play the opposite role in shifting the shock parameter space.

\subsubsection{Astrophysical Implications: Disk luminosity \& QPO frequency parameter space}

Corresponding to the $\varepsilon_{\rm in}$, $\lambda_{\rm in}$ parameter space presented in Fig.~\ref{fig_10}, we demonstrate the PS luminosity vs QPO frequency parameter space in Fig.~\ref{fig_11}. Due to the high temperature and density of the PS region, high energy radiation is expected to originate from this region. When $Q^{-}$ represents the cooling rate, $x_{\rm in}$ denotes the inner critical point, and $x_{\rm s}$ signifies the shock location, the PS surface disk luminosity ($L_{\rm PS}$) may be approximated as \citep{Luo2013ApJ,Sarkar2018},
$$
L_{\rm PS} = 4\pi \int_{x_{\rm in}}^{x_{\rm s}} x Q^{-} dx, 
\eqno(23)
$$
where we have assumed the mass of the BH as $M_{\rm BH}$ = 10 $M_{\odot}$. It is to be noted that in Fig.~\ref{fig_11}, $L_{\rm PS}$ is measured in Eddington units ($L_{\rm Edd}$). In the present paper, we follow \citet{2014A&AStraub,Yuan-Narayan2014,Sarkar2020RAA} and we use the definition of $L_{\rm Edd}$ given by $L_{\rm Edd}$ =  $\dot{M}_{\rm Edd} c^2$, where $\dot{M}_{\rm Edd}$ is the Eddington accretion rate given by $\dot{M}_{\rm Edd}$ = 1.39 $\times$ 10$^{17}$ $\times$ $M_{\rm BH}/M_{\odot}$ g s$^{-1}$ \citep{Sarkar2018,Das-Sarkar2018,Sarkar2020RAA}. The PS velocity profile is used to calculate the infall time $t_{\rm infall}$. If $u(x)$ is the PS velocity, $t_{\rm infall}$ is computed as \citep{Ryu1997ApJ,2014MNRASGarain,Aktar2017MNRAS,N2018ASSP},
$$
t_{\rm infall} = \int dt = \int_{x_{\rm s}}^{x_{\rm in}} \frac{dx}{u(x)}.
\eqno(24)
$$
Following \citet{1996ApJM,Aktar2017MNRAS,N2018ASSP}, the QPO frequency is calculated as follows:
$$
\nu_{\rm QPO} = \frac{1}{t_{\rm QPO}} \sim \frac{1}{t_{\rm infall}},
\eqno(25)
$$
in units of $\frac{r_{\rm g}}{c}$. After being multiplied by $\frac{c}{r_{\rm g}}$, the QPO frequency is converted to the Hertz unit. In each of the four cases demonstrated in Fig.~\ref{fig_11}, we find that for a given QPO frequency, a range of PS luminosity values is possible (depending on the other flow parameters), and the PS luminosity decreases with the increase of QPO frequency (for given $\Phi_{\rm s}$ and $m$ values). For a given QPO frequency and a fixed value of $\Phi_{\rm s}$, the PS luminosity is found to be generally higher when the wind parameter $m$ is higher. Since the accretion rate at the inner critical point is fixed while generating the parameter space, the flow has to start with a higher accretion rate at the outer edge when the wind parameter $m$ is increased. Since the accretion rate is higher in the flow, this can explain the higher PS luminosity exhibited.

\section{Summary and conclusions}                        

This paper comprehensively investigates the structure of magnetized shocked viscous AD surrounding a BH that is moderately rotating ($a_{\rm BH} = 0.5$) and considering TC and {\color{black}{TDWs}}. Here, we have mainly concentrated on the global characteristics of AFs with shock waves that originate from the existence of a centrifugal barrier close to the EH. The same collection of standard conservation equations describing the magnetized accreting flow that \citet{Das-Sarkar2018,Sarkar2020RAA} examined served as the foundation for our model. With this, we studied the impacts of TC, {\color{black}{TDWs}}, and $\lambda$ in the global accretion solutions and investigated their implications. We discuss our main findings below.

We have shown that the AF may traverse through more than one critical point, contingent upon the input parameters, and that such a flow may manifest a shock phenomenon. We obtained steady global solutions for transonic accretion around a rotating BH with a standing shock between the inner and outer critical points (Figs.~\ref{fig_3},~\ref{fig_4} \&~\ref{fig_7}). We demonstrated how variation in the SC constant ($\Phi_{\rm s}$) and wind parameter (${m}$) affect the dynamics of shocks. We find that the stationary shock front migrates away from the BH horizon with an increase in the $\Phi_{\rm s}$ and $m$ parameters for a moderately rotating BH ($a_{\rm BH} = 0.5$) and fixed injection parameters at the outer edge (Figs.~\ref{fig_4} \&~\ref{fig_7}). We provide the AD structure, averaged vertically, corresponding to the global shock solutions under consideration (Figs.~\ref{fig_5} \&~\ref{fig_8}). As $\Phi_{\rm s}$ grows, we show that, for given outer edge injection parameters, the flow's $\lambda$ increases (Fig.~\ref{fig_5}c) at a given radial coordinate $x$, while its radial infall velocity decreases (Fig.~\ref{fig_4}b), in keeping with the results of \citet{Faghei2012}. Moreover, as $\Phi_{\rm s}$ grows, we find that, for given outer edge injection parameters, the density of the AF rises and the temperature ($T_{\rm p}$) decreases, which are also consistent with the findings of \citet{{Wu2010SCPMA},Faghei2012}. Additionally, we note from Fig.~\ref{fig_6} that there are global accretion solutions with shock across a wide range of $\Phi_{\rm s}$$-$parameter and $\lambda$. Next, we examine how an increase in $m$$-$parameter affects the shock position in Fig.~\ref{fig_9}. We observe the existence of global shocked accretion solutions for an extensive range of $m$ and $\lambda$, given a particular set of injection parameters at the outer edge. Thus, shocked accretion solutions are not isolated occurrences in the AF but occur for a wide range of flow parameters. 

{\color{black}{For a relativistic flow with an adiabatic index $\gamma = 4/3$, the compression ratio $R$ $\sim$ 1 for weak shocks, while $R$ $\rightarrow$ 7 for the strongest shocks \citep{Das2001A&A}. Following \citet{2007MNRASDas}, $R$ $\sim$ 4 indicates strong shocks. \citet{Sarkar2020RAA} found stronger shocks closer to the BH horizon, with $R \sim 5$. Our results in Figs.~\ref{fig_6} and~\ref{fig_9} similarly show $R \sim 5$ for shocks near the BH horizon. For $\gamma = 5/3$, \citet{Frank2002} showed gas compression by a factor of 4 in strong shocks. Our findings agree with previous studies \citep{2007MNRASDas,Sarkar2016,Sarkar2018,Sarkar2020RAA}, indicating relatively strong shocks forming closer to the BH and decreasing shock strength ($\Theta$) and $R$ as the shock location recedes from the BH horizon.}}

{\color{black}{In this study, we adopt the treatment of \citet{Chakrabarti1989,CD2004,Sarkar-Das2018JApA} and model the shock as a thin, non-dissipative structure, as the energy flux is continuous across the shock front ($\varepsilon_+ = \varepsilon_-$). Thus, our model's computed luminosity originates solely from the rotational energy released in the PS region with no contribution of dissipated energy of the AF at the shock to the computed PS surface disk luminosity. It is noteworthy that the flow's heating rate, $Q^+$ (defined in equation 10), depends not only on the AF differential rotation ($d\Omega/dx$) but also on the momentum flux ($W$ + $\Sigma u^2$) of the AF. Corresponding to Figs.~\ref{fig_4}(a) and ~\ref{fig_7}(a), we have presented the radial variation of the logarithmic cooling rate ($Q^-$) in g/s$^3$ as a function of radial distance in Figs.~\ref{fig_5_addl}(a) and ~\ref{fig_8_addl}(a), respectively. The Fig.~\ref{fig_5_addl}(a) and Fig.~\ref{fig_8_addl}(a) show that $Q^-$ increases steadily as the AF approaches the BH horizon. Again, the momentum flux ($W$ + $\Sigma u^2$) is assumed continuous across the shock front, as discussed in Section 3.2. Following \citet{2007MNRASDas}, excess torque sources at the shock are neglected, ensuring angular momentum conservation across the shock front \citep{1995MNRAS_CM,CD2004,2018MNRASDDM}. As a result, both heating ($Q^+$) and cooling rates ($Q^-$) remain continuous across the shock. Thus, shock physics is inherently incorporated into PS surface disk luminosity calculations.}}

{\color{black}{We now discuss if TC predominantly acts as an energy sink or source in our model. TC plays a crucial role in supplementing heating within hot AFs, as demonstrated by \citet{Tanaka-Menou2006,2008MNRASA,Shadmehri2008}. Specifically, \citet{Tanaka-Menou2006} demonstrated that enhancing saturated conduction levels facilitates increased heat transfer from hotter, inner regions, leading to localized temperature enhancement in the gas relative to the classical advection-dominated AF (ADAF) solution. Following \citet{2008MNRASA,2013Ap&SSGhas}, we have considered that the energy generated by heat conduction into the volume and viscous dissipation is balanced by advection cooling. In our model, the energy produced by viscous heating has been proposed to be transported by saturated TC, considering the weakly collisional character of hot AFs. Thus, sufficient additional heating in ADs characterized by hot ADAFs has been proposed to be delivered by TC as a potential mechanism in our formalism.}}

{\color{black}{Next, we discuss some physical estimates of the $\Phi_{\rm s}$ parameter. According to the treatment of \citet{Cowie-McKee1977}, the dimensionless saturated conduction parameter, $\Phi_{\rm s}$, is smaller than unity. As reported by \citet{Tanaka-Menou2006}, their solutions match the conventional ADAF solutions for extremely small $\Phi_{\rm s}$. \citet{Tanaka-Menou2006} demonstrated by adding the $\Phi_{\rm s}$ that the influence of TC may be more clearly observed when we approach $\sim 0.001-0.01$. Consequently, \citet{2009MNRAS_Ghanbari} have looked at how TC works on ADAFs in this $\Phi_{\rm s}$ range. Further, several works in the literature have reported that even at extremely low values of the $\Phi_{\rm s}$ parameter, TC can substantially impact the evolution of ADAF systems \citep{Shadmehri2008,2008MNRASA,Faghei2012,2012bAp&SS_Faghei,2013Ap&SSGhas,2013MNRAS_Khajenabi,Rezgui2019ApJ,Rezgui2022MNRAS,2020MNRAS_Ghoreyshi,Mitra2023}. Notably, \citet{2013Ap&SSGhas} applied the model of hot magnetized AF, including TC on the astronomical object LMC X-3 with $M = 10 M_{\odot}$ for the two values $\Phi_{\rm s}$ = 0.004 and $\Phi_{\rm s}$ =  0.008 where the estimated observed flux and the continuum spectra of the source were found in agreement with earlier reported results. Again, by using the $\Phi_{\rm s}$ parameter range [0.002, 0.01], \citet{Rezgui2019ApJ} have adjusted the $\Phi_{\rm s}$ coefficient in their model to assess its sensitivity to the properties of AF around compact objects. Given these findings, we claim that the values of $\Phi_{\rm s}$ coefficient used in the present work lie in the physically motivated range, as reported in the literature for ADAF solutions.}}

{\color{black}{Moving forward, we examine the relative importance of various radiative processes within our formalism. Based on the treatment of \citet{Nar-Yi-1994}, we have used the parametric cooling prescription in our work, incorporating the constant cooling factor $f_c$. \citet{Yuan-Narayan2014} indicated that $f_c$ is nearly 1 for hot flows. The constant $f_c$ assumption in our model idealizes the complex radial dependence of $f_c$, which is governed by heating and cooling processes \citep{1998Narayan,2006ApJWatarai,2007PASJWatarai,2008MNRASA,2009MNRAS_Ghanbari,2009RAASinha,Faghei2012}. By setting $f_c = 0.8$, we effectively model an advection-dominated flow with suppressed radiative cooling, consistent with ADAF characteristics. In this scenario of weak cooling, according to the findings of \citet{1997ApJNara}, the details of the cooling processes when computing the AD dynamics are not required to be accounted for. The work by \citet{Aktar2017MNRAS}, using a parametric cooling prescription has shown the existence of steady shock solutions in the AF around spinning BHs for the entire range $0 \leq f_c \leq 1$ for $a_{\rm BH} = 0.5$. Moreover, as discussed in Section 2 (following Equation 11), bremsstrahlung, synchrotron, and inverse Comptonization significantly impact radiative cooling in AFs. However, bremsstrahlung emission is recognized as an inefficient cooling mechanism for stellar-mass BH systems \citep{2000IJMPD_CC,2007MNRASDas,Sarkar2018}, thus its contribution to energy loss may be neglected. Due to the compact size of stellar-mass BH ADs, magnetic fields substantially permeate the disk, suggesting synchrotron emission dominates over bremsstrahlung \citep{CC2002,RM2010b,Sarkar2020RAA}. Moreover, in our model, we assume a single-temperature plasma approximation, thereby we neglect the inverse Comptonization effect, which would require a more complicated two-temperature treatment \citep{Das-Sarkar2018,Sarkar2020RAA}.}} 

These results may have significant consequences. It has been observed that the steady shock solution vanishes when the $\Phi_{\rm s}$ and $m$ parameters lie beyond specific threshold values (see Figs.~\ref{fig_6} \&~\ref{fig_9}). According to \citet{Ryu1997ApJ} and \citet{Das2014MNRAS}, non-steady shock, however, may still manifest. In the meanwhile, it was demonstrated by \citet{1996ApJM} and \citet{CM2000ApJ} that the QPO frequency of the hard radiation that emerges from BHs is related to the time that accreting matter takes to fall into the BH. Using this prescription, we apply our formalism to determine how the PS luminosity ($L_{\rm PS}$) vs QPO frequency ($\nu_{\rm QPO}$) parameter space is modified as a function of $\Phi_{\rm s}$ and $m$ parameters. Our demonstration of the combined effect of $\Phi_{\rm s}$ and $m$ parameters on the formation of steady shocks in the form of ${\varepsilon}_{\rm in}-\lambda_{\rm in}$ parameter space (Fig.~\ref{fig_10}) and the corresponding $L_{\rm PS}-\nu_{\rm QPO}$ parameter space (Fig.~\ref{fig_11}), provides some bound on the AD luminosity and QPO frequencies from the shock model.

Another important conclusion from our investigation is worth highlighting. First, the oscillation of the shock is thought
to be the cause of the observed QPOs. Several BH candidates, including MAXI J1659-152, H 1743-322, and GX 339-4, have rapidly evolving QPOs during outbursts. During the declining phase of the outbursts, the QPO frequencies are monotonically falling from a few tens of Hz to a few mHz \citep{Nandi2012,2013AdSpRD}. According to \citet{Molla2016MNRAS,2013AdSpRD,2010A&AD,Nandi2012} and \citet{D2018ASSP}, this evolution of QPO frequency suggests that the shocks, in reality, propagate away from the BH horizon during the declining phase of the outbursts, as the QPO frequency is inversely proportional to the infall time of the PS flow. A reasonable explanation for such an occurrence may be found in our mechanism, which is detailed in the paper. Thus, the TC process and the generation of {\color{black}{TDWs}} seem to play an essential role during the declining phase of the outburst observed in BH candidates.

We finally want to discuss the shortcomings of the current formalism. To keep the analysis simple, we have used the \citet{ABN1996} PNP to simulate the general relativistic effects surrounding a rotating BH. The current formalism is not suitable for investigating the AF dynamics surrounding a maximally rotating BH since the results derived using the \citet{ABN1996} PNP deviate from the general relativistic findings by 10-20\%. In this context, it is worth mentioning that the PNP by \citet{Dihingia2018c}, although slightly complicated, is suitable to study the global accretion solutions around rapidly rotating BHs. Additionally, as the present paper aims to investigate the impact of $\Phi_{\rm s}$ and $m$ parameters on global shocked accretion solutions, we have consistently assumed $a_{\rm BH} = 0.5$ throughout our analysis, following a similar approach adopted by \citet{Saha2016}. Moreover, since we have considered thermal winds in our analysis, the role of magnetic fields and radiation driving in generating disk winds have not been taken into account in our work. The adiabatic index should be determined self-consistently from the AF's thermal properties, but it is treated as a global constant throughout the paper. Furthermore, for the sake of simplicity, we disregard the physically driven cooling mechanisms like bremsstrahlung, synchrotron, and Comptonization and instead focus on the parametric cooling mechanism. It is anticipated that the qualitative behavior of the results of this paper will remain the same due to the inclusion of these details. In the future, we want to address these matters.


\section*{Author statement}

The authors affirm that there was no conflict of interest throughout the manuscript preparation process.

\section*{Declaration of Competing Interest}

According to the authors, there is no indication that any of the results presented in this paper was impacted by recognised financial conflicts or personal ties.

\section*{Data availability}

No data was used for the research described in the article.



\section*{Acknowledgements}
BS acknowledges the financial support received from Tezpur University for this work (Office Order No- F.02-931/I(E)/1529 dated 10/08/2023). BS also acknowledges the financial support of UGC-BSR Start-Up-Grant for carrying out this work (No.F.30-558/2021 (BSR) dated-16/12/2021). In addition, BS would like to thank IUCAA for hosting him throughout his visits, where some of this work was completed, and he acknowledges the visiting associateship programme there. {\color{black}{We thank the reviewer for constructive feedback and suggestions which led to considerable improvement of the manuscript. BS also extends gratitude to Ankur Nath, Asish Jyoti Boruah, and Liza Devi for their valuable input and fruitful discussions on the manuscript.}}


\end{document}